\documentclass[usegraphicx,usenatbib]{mn2e}
\usepackage[usenames,dvipsnames]{xcolor}

\usepackage{amssymb}
\usepackage{epsf,multirow}
\usepackage{graphicx}
\usepackage{txfonts}
\usepackage{ulem}

\def\aap{A\&A}
\def\apj{ApJ}
\def\apjs{ApJS}
\def\apjl{ApJL}
\def\apss{Ap\&SS}
\def\mnras{MNRAS}
\def\aj{AJ}

\def\araa{ARA\&A}   

\def\Ks{$K_{\rm short}$}
\def\Kl{$K_{\rm long}$}
\def\kms{~km~s$^{-1}$}
\def\micron{~$\mu$m}

\def\deg{$^{\circ}$}

\def\Dv{\Delta{\rm v}}
\def\n{NGC\,}

\def\H2{H$_2$}

\def\PII{[P\,{\sc ii}]}
\def\AlIX{[Al\,{\sc ix}]}
\def\CaVIII{[Ca\,{\sc viii}]}

\def\OIII{[O\,{\sc iii}]}

\def\NeV{[Ne\,{\sc v}]}
\def\FeII{[Fe\,{\sc ii}]}
\def\FeVII{[Fe\,{\sc vii}]}
\def\FeX{[Fe\,{\sc x}]}
\def\FeXI{[Fe\,{\sc xi}]}

\def\SiVI{[Si\,{\sc vi}]}
\def\SiVII{[Si\,{\sc vii}]}

\def\SIX{[S\,{\sc ix}]}
\def\SXII{[S\,{\sc xii}]}

\def\PaB{Pa$\beta$}
\def\BrG{Br$\gamma$}

\def\mnras{MNRAS}
\def\apj{ApJ}

\title[The CLR of \n1068]{Resolving the coronal line region of \n1068 with near-infrared integral field spectroscopy\thanks{Based on observations obtained at the Gemini Observatory, which is operated by the Association of Universities for Research in Astronomy, Inc., under a cooperative agreement with the NSF on behalf of the Gemini partnership: the National Science Foundation (United States), the Science and Technology Facilities Council (United Kingdom), the National Research Council (Canada), CONICYT (Chile), the Australian Research Council (Australia), Minist\'erio da Ci\^encia, Tecnologia e Inova\c{c}\~{a}o (Brazil) and Ministerio de Ciencia, Tecnolog\'ia e Innovaci\'on Productiva (Argentina).}}

\author[Mazzalay et al.]{X. Mazzalay$^{1}$\thanks{E-mail: ximena@mpe.mpg.de}, A. Rodr\'iguez-Ardila$^2$, S. Komossa$^3$ and Peter J. McGregor$^4$\\
$^{1}$Max-Planck-Institut f\"ur extraterrestrische Physik, Postfach 1312, 85741 Garching, Germany\\
$^2$Laborat\'orio Nacional de Astrof\'isica, Rua Estados Unidos 154, 37504-364 Itajub\'a, MG, Brazil\\
$^3$Max-Planck-Institut f\"ur Radioastronomie, Auf dem H\"ugel 69, 53121 Bonn, Germany\\
$^4$Research School of Astronomy and Astrophysics, The Australian National University, Cotter Road, Weston, ACT 2611
}

\begin{document}

\date{}


\maketitle


\begin{abstract}

We present adaptive optics-assisted $J$- and $K$-band integral field spectroscopy of the inner $300\times 300$~pc of the Seyfert~2 galaxy \n1068. The data were obtained with the Gemini NIFS integral field unit spectrometer, which provided us with high-spatial and high-spectral resolution sampling. The wavelength range covered by the observations allowed us to study the \CaVIII, \SiVI, \SiVII, \AlIX\ and \SIX\ coronal line (CL) emission, covering ionization potentials up to 328~eV.
The observations reveal very rich and complex structures, both in terms of velocity fields and emission-line ratios.
The CL emission is elongated along the NE-SW direction, with the stronger emission preferentially localized to the NE of the nucleus. CLs are emitted by gas covering a wide range of velocities, with maximum blueshifts/redshifts of $\sim -1600/1000$\kms.
There is a trend for the gas located on the NE side of the nucleus to be  blueshifted while the gas located towards the SW is redshifted.
The morphology and the kinematics of the near-infrared CLs are in very good agreement with the ones displayed by low-ionization lines and optical CLs, suggesting a common origin.
The line flux distributions, velocity maps, ionization structure (traced by the \SiVII/\SiVI\ emission-line ratio) and low-ionization emission-line ratios (i.e. \FeII/\PaB\ and \FeII/\PII) suggest that the radio jet plays an important role in the structure of the CL region of this object, and possibly in its kinematics.

\end{abstract}


%

\section{Introduction}

Feedback due to outflows and jets is believed to be a major driver of the co-evolution of galaxies and their supermassive black holes \citep[review by][]{Fabian12}. Radio jets interact in various ways with their surroundings, sweeping up, and driving shocks into the ambient gas. While direct jet-gas interaction locally affects the velocity field of the surrounding medium, shocks or the radiation field from fast, auto-ionizing shocks can significantly influence or dominate emission-line strengths. Nearby galaxies are key laboratories for studying jet-gas interaction in great detail, and at high resolution, through spatially resolved spectroscopy of the (high-ionization) emission lines.

A subset of active galactic nuclei (AGN) show in their spectra emission lines from very highly ionized atoms, known as `coronal lines' (CLs). They are collisionally excited forbidden transitions within low-lying levels of ionized species with ionization potentials IP $> 100$~eV. CLs have been detected in the optical and infrared spectra of all types of AGN, including Seyfert 1 and Seyfert 2 galaxies, narrow-line Seyfert 1 galaxies, and radio galaxies \citep[e.g.][]{Penston84, Marconi94, Nagao00, Sturm02, Rodriguez-Ardila02a, Rodriguez-Ardila06b, Deo07, Mullaney08, Komossa08, Gelbord09}. Their abundance is approximately similar in AGN of type 1 and 2 \citep{Rodriguez-Ardila11} and they represent one of the major gaseous components of active nuclei.

The precise nature and origin of CLs are still a matter of debate. Different scenarios have been considered in the literature, including winds from the molecular torus \citep[e.g.][]{Pier95, Nagao00, Mullaney09}, (X-ray) ionized absorbers \citep[NLR; e.g.][]{Komossa97a, Komossa97b, Porquet99}, a high-ionization component of the inner narrow-line region \citep[e.g.][]{Komossa97c, Ferguson97b, Binette97}, and a low-density component of the interstellar medium \citep[ISM;][]{Korista89}. However, up to now none of the above hypotheses has been fully discarded because of the lack of high-spatial resolution observations on a sizeable number of targets.

The mechanisms powering the CLs have also been a subject of strong debate, with two competing processes usually invoked: photoionization from the central engine and shocks between the jets and the NLR gas. \citet{Dopita96} and \citet{Bicknell98} proposed that the emission in the NLR may be entirely caused by shocks. Velocity splitting of over 1000\kms, reported by \citet{Axon98}, in the vicinity of some of the bright emission-line knots in the NLR of \n1068 provides evidence of the existence of fast shocks in AGNs. However, Hubble Space Telescope (HST) data \citep{Crenshaw00a, Cecil02, Mazzalay10} have shown emission-line ratios consistent with photoionization instead of shock heating mechanisms. Moreover, models accounting for both photoionization from the central radiation source and shocks were required to explain both the continuum and NLR spectrum, including the CLs, observed in the spectra of a sample active galaxies \citep{Contini03, Rodriguez-Ardila06b}. Furthermore, sometimes the same data sets have been explained by different scenarios. In particular, both mechanisms, shocks and photoionization, may plausibly be at work in the same galaxy. Therefore, high-resolution studies of nearby galaxies are essential in order to disentangle the effects and contributions of shocks and photoionization.

Recent advances in the study of CLs include the determination of the size and morphology of the region emitting these lines by means of optical HST and ground-based near-infrared (NIR) adaptive optics (AO) imaging/spectroscopy in a few AGNs \citep[e.g.][]{Riffel08, Storchi-Bergmann09, Mazzalay10, Muller-Sanchez11}. \citet{Mazzalay10},
for instance, presented an analysis of the optical CLs of a sample of 10 Seyfert galaxies, including \n1068. These authors used the excellent spatial resolution provided by STIS/HST to resolve the CL region (CLR) and study the properties of the CLs as a function of the distance to the nucleus. The optical CLs were found to be very complex, showing strong variations in their intensities and profiles along the spatial direction, which in most cases was nearly aligned to the position of the radio jet. These lines are emitted from very compact regions ($\sim 30$~pc) up to regions of $\sim 250$~pc. The morphology and the kinematics of the lower-ionization CLs (i.e. \NeV, \FeVII\ and \FeX) are very similar to that of lower ionization lines such \OIII. On the other hand, CLs with IP $> 250$~eV (i.e. \FeXI\ and \SXII) are emitted in much more compact regions and do not seem to share the same overall kinematics found for the lower ionization lines. Several pieces of evidence (e.g. lack of correlation between the radio and CL emission,  and relatively good success of photoionization models in reproducing measured line ratios) point towards photoionization from a central source as the main mechanism of formation of CLs.

However, the STIS results are restricted by the relatively small region covered by the slit width and along a particular direction that can be analysed with long-slit spectroscopy. In spite of the typical 0.1~arcsec spatial resolution, the limited area covered by the slit did not allow rejection of jet-NLR gas interaction in the production of the CLs \citep{Mazzalay10}. A further step on this subject is expected to be achieved by means of AO integral field spectroscopy, which can give us a more detailed view of the innermost parsecs of nearby AGNs at similar angular resolutions as HST.

\n1068 is a natural candidate for a follow up with AO integral field spectroscopy. This object is one of the nearest and probably the most intensely studied Seyfert 2 galaxy in the literature. Observations in all wavelength bands from radio to hard X-rays have formed a uniquely detailed picture of that source. It hosts a prominent NLR that is approximately co-spatial with a linear radio source with two lobes \citep{Wilson83}, extensively characterized from subarcsecond clouds probed by the HST \citep{Evans91, Macchetto94} to the ionization cone and extended emission-line region \citep{Pogge88, Unger92} extending to radii of at least 30~arcsec ($1~{\rm arcsec}=73$~pc for $z=0.003793$ and ${\rm H_o} = 75$\kms~Mpc$^{-1}$).

The optical and NIR CLs of \n1068 were first studied by \citet{Marconi96}. They found that the flux distribution peaks 50~pc NE of the nucleus and appears to arise predominantly in outflowing gas within the prominent ionization cone aligned with the radio jet. The line ratios argue against collisional ionization but could be explained with photoionization by radiation from the active nucleus. \citet{Prieto05} presented a \SiVII~2.48\micron\ narrow band image obtained with the IR camera NACO at the ESO/VLT (spatial resolution ${\rm FWHM}=6.8$~pc or 0.097~arcsec). The images show a diffuse coronal region with filamentary structure. The extent of the CL emission is larger than that predicted by photoionization models, which argues for additional in situ gas excitation, the most plausible energy source being shock excitation. \citet{Rodriguez-Ardila06b}  reported a double-peak structure in the [Fe\,{\sc vii}]~$\lambda$6087 line, with the relative separation between the two peaks increasing with the distance from the centre, signalling the presence of a highly energetic outflow, not detected in lower ionization lines. Very recently, \citet[hereafter MS11]{Muller-Sanchez11} analysed the morphology and kinematics of \SiVI~1.96\micron\ (IP=166~eV) CL of \n1068 by means of $K$-band AO integral field spectroscopy SINFONI data. These authors confirmed the great complexity displayed by the \SiVI\ emission-line region of this galaxy and favoured a scenario where the radio jet has no strong influence in the formation and kinematics of the CLR, probably only affecting the kinematics of some emission-line knots located at distances of less than 1~arcsec from the nucleus, towards the NE.

Because of the above points, additional information is needed to remove these ambiguities and identify the dominant mechanism powering the CLs.
Moreover, the role of the radio jet in shaping the morphology and kinematics of the inner NLR gas can now be studied in greater detail thanks to the availability of spatial resolution at the parsec scale using ground-based facilities. Here, we present AO-assisted $J$- and $K$-band integral field spectroscopy data of the $300 \times 300$ central parsecs of \n1068 obtained with NIFS at Gemini North. The wavelength range covered by these data (1.15--2.5~\micron) allowed us to study in a spatially resolved fashion the \CaVIII, \SiVI, \SiVII, \AlIX\ and \SIX\ CL emission,  covering in this way ionization potentials up to 328~eV.
In Section~\ref{s_obs}, we describe the observations and data reduction procedures. The analysis of the morphology and kinematics of the CL gas is described in Sections~\ref{s_morphology} and \ref{s_kinematics}, respectively. Section~\ref{s_lineratios} describes the ionization structure of the CLR and
the analysis of the origin of the \FeII\ emission in this galaxy. The summary and final conclusions can be found in Section~\ref{s_summary}.

\section{Observations and data reduction}\label{s_obs}

Spectra of \n1068 were obtained with NIFS \citep[][]{McGregor03} on the Gemini North telescope in December 2006 under programme GN-2006B-C-9. The observations and data reduction have been described in detail by \citet{Storchi-Bergmann12}. NIFS was used with the ALTAIR facility adaptive-optics system in its natural guide star mode. The compact Seyfert nucleus of \n1068 was used as the adaptive-optics reference object. The $J$, \Ks\ and \Kl\ observations discussed in this paper were obtained on 2006 December 15, 12 and 13 (UT), respectively. The $J$-band spectra have a resolving power of $\sim 6040$, and the $K$-band spectra have a resolving power of $\sim 5290$.

NIFS is an image-slicer integral-field spectrograph with a square field of view of $3\times 3~{\rm arcsec}^2$, which is divided into 29 slitlets each $0.103$~arcsec wide with a sampling of $0.044$~arcsec  along each slitlet. The instrument was set to a position angle on the sky of 300\degr\ to align the slitlets approximately perpendicular to the axis of the radio jet and ionization cone observed in the nucleus of \n1068. This provided coarser spatial sampling along the jet axis, and finer spatial sampling across the jet.

Each data set consisted of sets of multiple 90~s exposures that began with an offset sky position and were followed by nine galaxy exposures obtained in a $3 \times 3$-frame grid centred on the galaxy nucleus. A frame offset of $1.0$~arcsec was applied along and perpendicular to the NIFS slitlets, so that the full field of view is $5.0 \times 5.0~{\rm arcsec}^2$ on the sky, with maximum exposure in the central $3.0 \times 3.0~{\rm arcsec}^2$. This exposure grid was repeated nine times for each grating setting. The nucleus of \n1068 was within each frame, so it was used to spatially register the exposures. An arc lamp exposure and a measurement of a telluric standard star were obtained after each data set.

The data reduction was performed using the {\sc gemini iraf nifs} package and followed standard procedures as described by \citet{Storchi-Bergmann12}. The data were flux calibrated using the observations of the telluric standard star, HIP 5886, which was assumed to have the intrinsic spectrum with a level set by its 2MASS magnitude in the wavelength band of the spectrum and a shape defined by a 8600~K blackbody, which has the same 2MASS $J-K$ colour as HIP 5886. The data reduction resulted in single data cubes at each grating setting with all instrumental signatures removed.

\section{Results and discussion}\label{s_results}

\subsection{The CL spectrum of \n1068}\label{s_nirspectrum}

\begin{figure*}
\centering
\includegraphics[width=12cm]{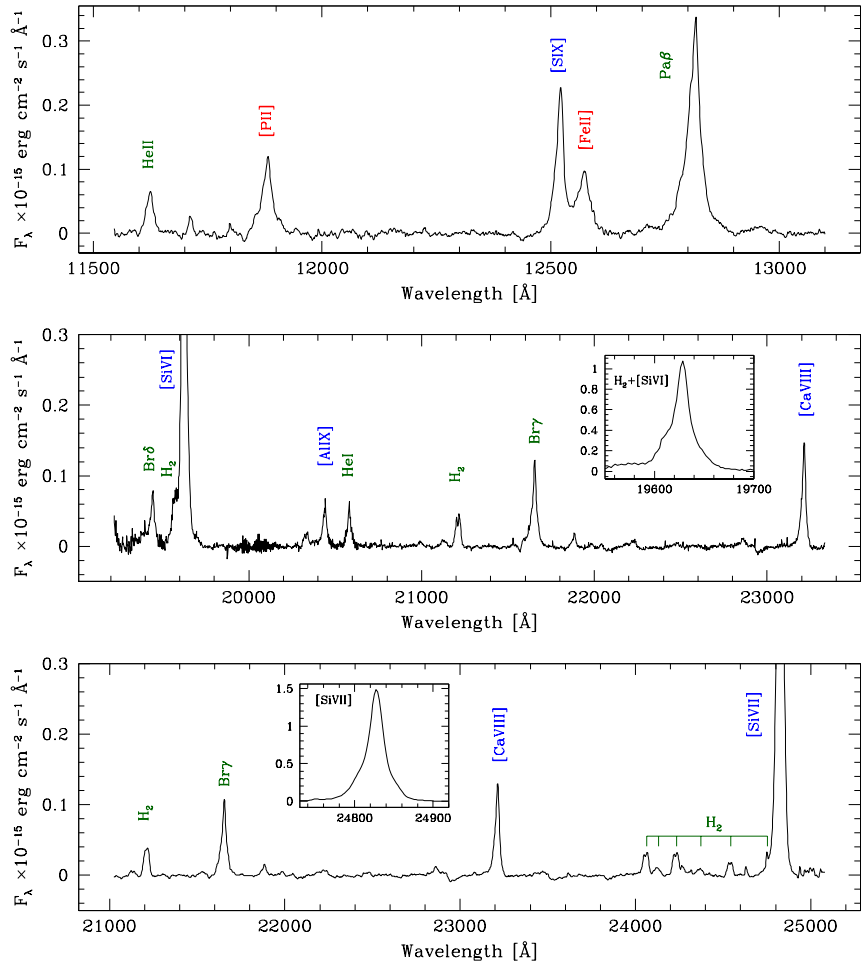}
\caption {\n1068 continuum-subtracted spectra in the rest frame of the $J$ (upper panel), $K_{\rm short}$ (middle panel) and $K_{\rm long}$ (lower panel) bands extracted from a $\sim 0.2 \times 0.2~{\rm arcsec}^2$ region centred at $\sim 0.6$~arcsec north of the nucleus. The labels mark the position of the principal  emission lines observed in the spectra.}
\label{f_JKspectra}
\end{figure*} 

The NIR spectrum of \n1068 displays numerous emission lines, the brightest of which are \PaB, \FeII~1.26\micron, several \H2\ lines, and the high-ionization CLs \SIX~1.25\micron\ (IP=328~eV), \SiVII~2.48\micron\ (IP=205~eV) and \SiVI~1.96\micron\ (IP=166~eV). Additionally, \n1068 exhibits weak CL emission from \AlIX\ at 2.04\micron\ (IP=285~eV) and \CaVIII\ at 2.32\micron\ (IP=128~eV).
Fig.~\ref{f_JKspectra} shows the $J$, $K_{\rm short}$ and $K_{\rm long}$ continuum-subtracted spectrum of \n1068 integrated in a region size of $\sim 0.2\times 0.2~{\rm arcsec}^2$ located at a distance of $\sim 0.6$~arcsec north from the nucleus. It can be seen that the CLs virtually dominate the NIR spectrum. The strongest CLs observed, \SIX, \SiVI\ and \SiVII, are blended with the \FeII~1.26\micron, \H2~1.96\micron\ and \H2~2.48\micron\ emission lines, respectively. Therefore, in order to study the CL emission it was first necessary to subtract the contribution from the \FeII\ and \H2\ lines. For this purpose, we first used the {\sc continuum} task of {\sc iraf} to subtract the continuum emission from the spectra of the data cubes.
The simplest case to handle was the \SIX$+$\FeII\ blend, since these two lines are not very strongly mixed. For each spectrum, we represented the profile of each of these lines with a Gaussian function, and subtracted the \FeII\ emission from the data. 
This procedure was satisfactory for most of the spectra, although in some regions it was necessary to include an extra Gaussian component to account for the strong asymmetries or double peaks displayed by the \FeII\ line. These regions are located towards the SW of the nucleus, where the \FeII\ line splits into two components, generally one redshifted and one blueshifted (up to $\sim -450$~\kms). Note that, even though the \FeII\ show large blueshifts, the separation in wavelength with the \SIX\ line is still large enough to allow a clear deblending of the lines. Detailed analysis of the kinematics of the \FeII\ emitting gas, based on this same data set, can be found in Riffel et al. (in preparation).
Deblending the \SiVI$+$\H2\ lines was more challenging than the previous case because of the strong broad and complex profile displayed by \SiVI\ in many regions, making the discrimination of each line difficult (see Fig.~\ref{f_JKspectra}). For this task, we used as reference for \H2~1.96\micron\ the \H2~2.12\micron\ line. The latter was chosen because it is located relatively close in wavelength to the former and in a region free of any other features. We modelled the \H2~2.12\micron\ line as the sum of two Gaussian components (to account for the asymmetries or double peaks displayed by this line in some regions), and assigned the same widths, velocity shifts and relative peak intensities of these two components to the ones used to describe the emission of \H2~1.96\micron. Additional constraints were necessary in some critical regions where the width and the intensity of the \SiVI\ line did not allow a clear discrimination between the coronal and molecular lines. In these regions, we applied an additional restriction, fixing the value of the \H2~1.96\micron/2.12\micron\ line ratio to 1.08. This corresponds to the mean value derived from the measurements in the regions where the \H2$+$\SiVI\ blend was not too strong. Although this line ratio is not independent of the physical properties of the emission-line gas, the measured values showed no significant variations between the spectra.
Once modelled satisfactorily, the \H2~1.96\micron\ emission was subtracted from the spectra.

As for the \SiVI\ line above, the observed \SiVII\ line is located in a spectral region containing \H2\ emission lines.
Nevertheless, because of the low intensity and particular distribution of the \H2\ emission (see Section~\ref{s_morphology}), its contribution to the \SiVII\ emission is practically negligible. Only when the \SiVII\ is analysed in velocity bins (Section~\ref{s_kinematics}), a clear contamination is present in the region where the \H2\ emission reaches its maximum ($\sim 1$~arcsec east of the nucleus). Since the \H2\ emission does not contribute significantly to the \SiVII\ emission and, when it does, it can be easily pinpointed, we did not subtracted the \H2\ emission lines from the \Kl\ NIFS spectra.

\subsection{Morphology and extension of the CLR}\label{s_morphology}

\begin{figure*}
\centering 
\includegraphics[width=12cm]{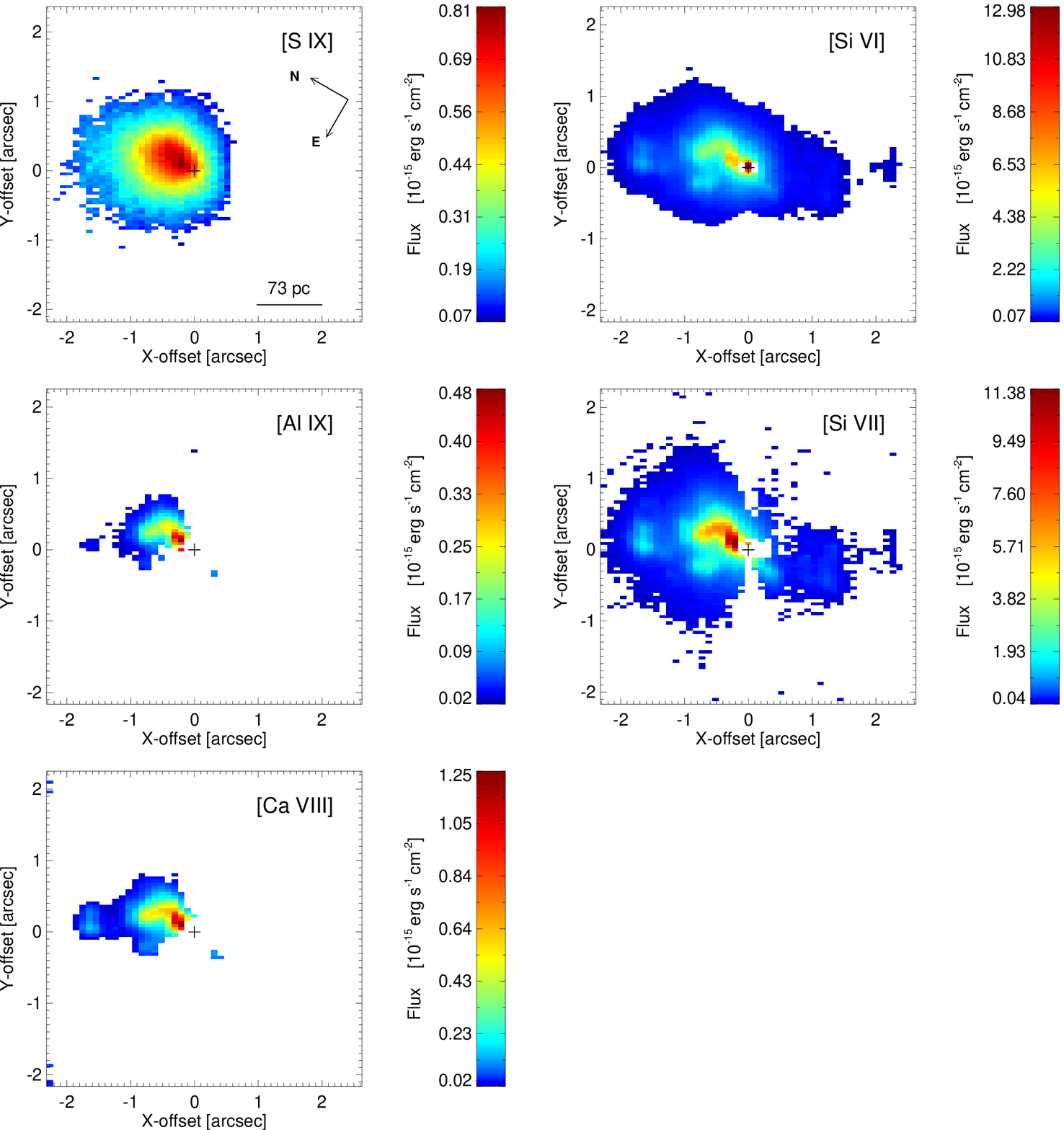}
\caption {Flux distribution maps of the CLs observed in the $J$ and $K$ data cubes. Each line is indicated in the upper right corner of the corresponding panel. The colour bars indicate the values of integrated flux in units of $10^{-15}$~erg~cm$^{-2}$~sec$^{-1}$. The orientation and spatial scale (indicated in the upper left panel) are the same for all the panels. The position of the continuum emission peak is indicated in each panel by a plus symbol.}
\label{f_mapasCLs}
\end{figure*}

\begin{figure*}
\centering
\includegraphics[width=12cm]{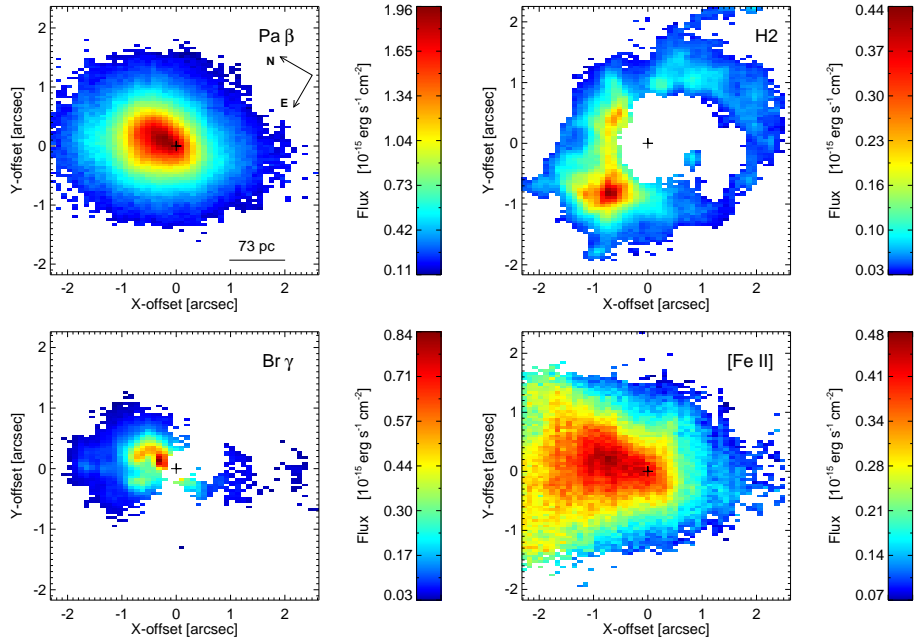}
\caption {Flux distribution maps of \PaB, \H2~2.12\micron, \BrG, and \FeII~1.26\micron. The colour bars indicate the values of integrated flux in units of $10^{-15}$~erg~cm$^{-2}$~sec$^{-1}$. The orientation and spatial scale (indicated in the upper left panel) are the same for all the panels. The position of the continuum emission peak is indicated in each panel by a plus symbol.}
\label{f_mapasLIL}
\end{figure*}

\begin{figure*}
\centering
\includegraphics[width=12cm]{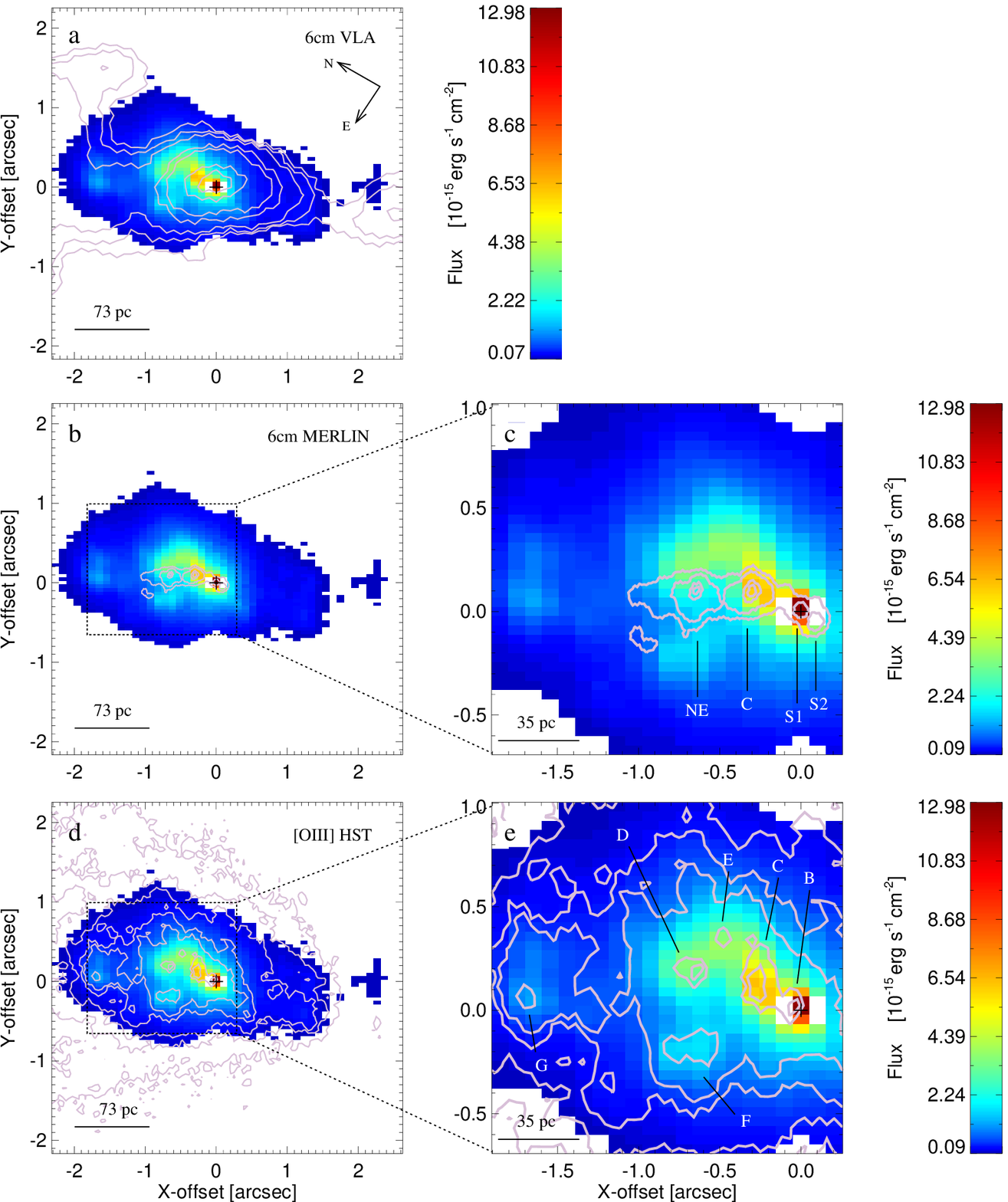}
\caption {Flux distribution maps of \SiVI, in units of $10^{-15}$~erg~cm$^{-2}$~sec$^{-1}$. The contours correspond to the 6~cm emission observed with VLA (panel a) and MERLIN (panels b and c) and \OIII\ emission (panels d and e). The labels S1, S2, C and NE indicated in panel c correspond to the radio knots defined by \citet{Gallimore96}. The labels B--G indicated in panel e correspond to the emitting clouds identified by \citet{Evans91}.}
\label{f_mapascontornos}
\end{figure*}

After applying the cleaning process described above, we constructed bi-dimensional maps of the flux distribution of the CLs observed in the $J$, \Ks, and \Kl\ data cubes of \n1068. We also constructed flux distribution maps of other lines of interest, namely \PaB, \BrG, \H2~2.12\micron, and \FeII~1.26\micron, for comparison purposes. The maps were constructed by integrating the flux under the emission-line profiles observed in the continuum-subtracted spectra. We used wavelength bins wide enough to include the emission from the highest velocity line-emitting gas. Figs.~\ref{f_mapasCLs} and \ref{f_mapasLIL} show the flux distribution maps of the CLs and the selected low-ionization and molecular lines of the inner $\sim 5\times 5~{\rm arcsec}^2$ of \n1068. The maps are centred at the position of the continuum peak of the corresponding $J$- and $K$-band, which coincides with the location of the central black hole \citep{Thompson01}. In these maps, we only report values with a signal-to-noise ratio (SNR) higher than 2.

Visual inspection of Fig.~\ref{f_mapasCLs} shows that the region emitting the CLs of \n1068 is highly inhomogeneous and elongated along the NE-SW direction. The inhomogeneous distribution can also be seen in the flux distribution maps of the low-ionization lines (i.e. \PaB, \BrG\ and \FeII, Fig.~\ref{f_mapasLIL}). The \SiVI\ map presented here is in agreement with the one reported by MS11. However, some differences are observed due to the fact that these authors did not subtract the \H2~1.96\micron\ emission, which can severely contaminate the \SiVI\ map at some particular locations. Although the integrated flux distribution is not strongly affected by this, the analysis of the velocity channels is more compromised, especially in the high-velocity blueshifted bins (see Section~\ref{s_kinematics} for details).
The extension of the \SIX, \SiVI\ and \SiVII\ CLs towards the NE from the nucleus is very similar, extending to a distance of 2.3~arcsec ($\sim 170$~pc). Note, however, that the NIFS data do not cover regions beyond this. Therefore, we cannot exclude the possibility of emission at larger distances from the nucleus. \CaVIII\ and \AlIX\ are observed in a more compact region, extending up to $\sim 140$~pc NE from the nucleus. Towards the SW the CLs are weaker, displaying an abrupt decrease in their intensity. \SiVI\ and \SiVII\ show a secondary peak of emission at $\sim 2.2$~arcsec SW, which can also be seen in the lower-ionization lines of \PaB, \BrG\ and \FeII\ (Fig.~\ref{f_mapasLIL}). This feature is also present in the data presented by MS11, confirming that it is real and not an instrumental artefact. Although the extension of \SIX\ towards the NE is very similar to that of \SiVI\ and \SiVII, towards the SW this line is observed only up to a few arcseconds from the nucleus, and no emission is detected at the position of the emission peak observed at $\sim 2.2$~arcsec SW in the other CLs. Note that, when comparing the spatial extension of different lines one must keep in mind the SNR at which these lines are detected, as the different sizes could be a simple artefact of the lower SNR of the weaker lines.

At this point, it is useful to compare the CL emission found here with the continuum and line distributions at other spectral regions. \n1068 harbours a nuclear jet observed in radio, which could be playing an important role in the ionization structure and kinematics of the NLR and, in particular, of the CLs of this galaxy \citep[e.g.][]{Macchetto94, Axon98, Mazzalay10}. Panels a-c of Fig.~\ref{f_mapascontornos} show a superposition of the \SiVI\ CL map and the 6~cm radio emission observed at high- and low-resolutions with VLA and MERLIN, respectively. The radio data were taken from the work of \citet{Gallimore96}. Moreover, it is possible to compare the infrared CL emission with that of the optical \OIII~5007~\AA\ line at similar spatial resolutions by means of the F502N WFPC2/HST image \citep{Dressel97}. The \OIII\ image was taken from the HST archive and its contours are shown in panels d and e of Fig.~\ref{f_mapascontornos}, superimposed on the \SiVI\ map. For the superposition of the different data sets we assumed that the maximum of the NIR continuum and the S1 knot of the jet observed with MERLIN coincide with the location of the central black hole, while the \OIII\ B cloud is slightly offset towards the NE \citep[see discussion in][and references therein]{Galliano03}.

It is easy to see from Figs.~\ref{f_mapasCLs} and \ref{f_mapascontornos} that the regions where the CLs are more intense are the ones located towards the N-NE from the nucleus, forming an arc that starts in the nucleus and extends about 0.8~arcsec towards the NE. This arc is resolved into individuals clouds by the HST \OIII\ image (clouds B, C, E and D of panel e of Fig.~\ref{f_mapascontornos}). Noteworthy is the spatial coincidence between the \OIII\ and CLs emission, both in large and small scales. The CLs peak at $\sim 1.7$~arcsec NE can be associated with the cloud G of \OIII, while cloud F also has a counterpart of CL emission except for \AlIX, whose emission is not detected in this region. The only exception to this point-to-point correspondence is the secondary peak of emission displayed by the \SiVI\ and \SiVI\ lines at 2.2~arcsec towards the SW, which is not observed in the \OIII\ emission. The lack of detection of this latter line in the optical is probably due to the strong extinction affecting \n1068 in the SW \citep[][and references therein]{Mazzalay10}.

An inspection of the \SIX\ map in Fig.~\ref{f_mapasCLs} shows that the spatial distribution of this line is smoother and more symmetric than the other CLs. This can be explained by the poorer performance of the AO system at short wavelengths, which results in a loss in spatial resolution of the $J$-band data compared to ones at the $K$-bands.
The comparison of the flux distribution maps of \PaB\ and \BrG\ (Fig.~\ref{f_mapasLIL}) confirms this hypothesis: they are formed under the same physical conditions and whenever both are detected, they should exhibit a similar distribution (albeit a scaling factor). However, it is clear from Fig.~\ref{f_mapasLIL} that \BrG\ traces more clearly the NLR structure of this galaxy, while the \PaB\ distribution is more diffuse.

The excellent match between the flux distribution of the NIR CLs and the optical \OIII\ emission together with the similar kinematics displayed by the optical low- and high-ionization lines \citep{Mazzalay10} suggest that the CLs and the low-ionization lines are produced by the same set of NLR clouds, or, else, that the NLR is clumpy on small scales ($< 15$~pc), with a mix of clouds at high and low density. If we assume that the former is the case, then we can set an upper value for the density of the CL gas as it must have densities lower than the critical density of the \OIII~5007~\AA, $n_c=7\times10^5$~cm$^{-3}$. That is, much lower than the critical density of the CLs studied here ($> 10^8$~cm$^{-3}$). This value is in complete agreement with the density of the CLR found to be between 10$^{5}$ and 10$^{6}$~cm$^{-3}$ by Rodr\'{\i}guez-Ardila et al. (in preparation), based on the analysis of optical [Fe\,{\sc vii}] flux ratios measured for a sample of 12 well-known AGNs.

The emission of both low- and high-ionization lines (including the optical \OIII~5007~\AA) are elongated along the direction of the jet observed in 6~cm with VLA, with an opening angle slightly larger than the one displayed by the jet (panel a of Fig.~\ref{f_mapascontornos}). At the smaller scale structures traced by MERLIN (panels b and c of Fig.~\ref{f_mapascontornos}) the strongest NIR CL emission, particularly that of \SiVI\ and \SiVII, coincides with the first portion of the jet, from S1 to C. At point C, the jet bends from its original direction and regions of strong CL  emission are now prominent at the NW and SE edges of the jet. This is compatible with the scenario suggested by \citet{Macchetto94} based on the \OIII\ images, where the jet cleans a channel trough the ISM, leaving behind hot ionized gas. As a result, the line emission is boosted along the edges of the jet where the gas is compressed. Additionally, cloud G is located in the region where the collimated radio jet becomes lobe-like.

Although the 6~cm radio emission is also observed towards the SW with VLA, the small-scale observations of MERLIN do not show evidence of emission in that direction. \citet{Pecontal97} convolved the MERLIN data in order to obtain a spatial resolution comparable to that obtained with VLA, but did not detect emission. This suggests that the MERLIN observations are not deep enough to detect the SW jet structures. Therefore, it is not possible to make a small-scale comparison between the jet and the CL emission in these regions. Nevertheless, the spatial relation between the radio jet and the CL emission is remarkable, suggesting that both are closely related.
Indeed, very recently, \citet{Wang12} presented exquisite Chandra X-ray images of the inner few hundred parsecs of \n1068 at a spatial resolution matching that of NIFS data. Extended X-ray emission is seen following that of \OIII\ HST, the CL emission (e.g. \SiVI) and Merlin 6~cm. They found strong evidence of shocks, possibly related to the radio jet affecting knots B$+$C, cloud G and the region located 1.2~arcsec SW of the nucleus while the jet thrusts through clouds D$+$E and F without sings of strong interaction. This issue is explored further in Section~\ref{s_lineratios}.

\subsubsection{Comments on the \H2\ morphology}

The \H2\ map (Fig.~\ref{f_mapasLIL}) shows a rather different distribution from those of the other lines, displaying a ring-like structure around the nucleus, with the strongest emission located along the E-W direction at a distance of $\sim 0.7$~arcsec towards the NE. The same distribution is displayed by other \H2\ lines observed in \n1068, here we show the \H2~2.12\micron\ line as a example. The \H2\ emission in \n1068 has been studied in detail by \citet{Galliano02}. These authors suggested that it traces a rotating molecular disc, consistent with the interferometric observations of the $^{12}$CO(2--1) reported by \citet{Schinnerer00}.
Additionally, \citet{Muller-Sanchez09} presented a study of the \H2\ emission of \n1068 from SINFONI/VLT integral field spectroscopy, focused on the central 1~arcsec. They interpret the data as evidence of hot gas, on scales of a few parsecs, fuelling the AGN.
The data presented here are consistent with that of \citet{Galliano02} and \citet[see their fig.~1]{Muller-Sanchez09}, regarding the size, shape and orientation of the \H2\ distribution \citep[see also][]{Storchi-Bergmann12}. A more detailed analysis of the molecular gas using this same data set is carried out by Riffel et al. (in preparation).

\subsection{Kinematics of the CLs}\label{s_kinematics}

\begin{figure*}
\centering
\includegraphics[width=12cm]{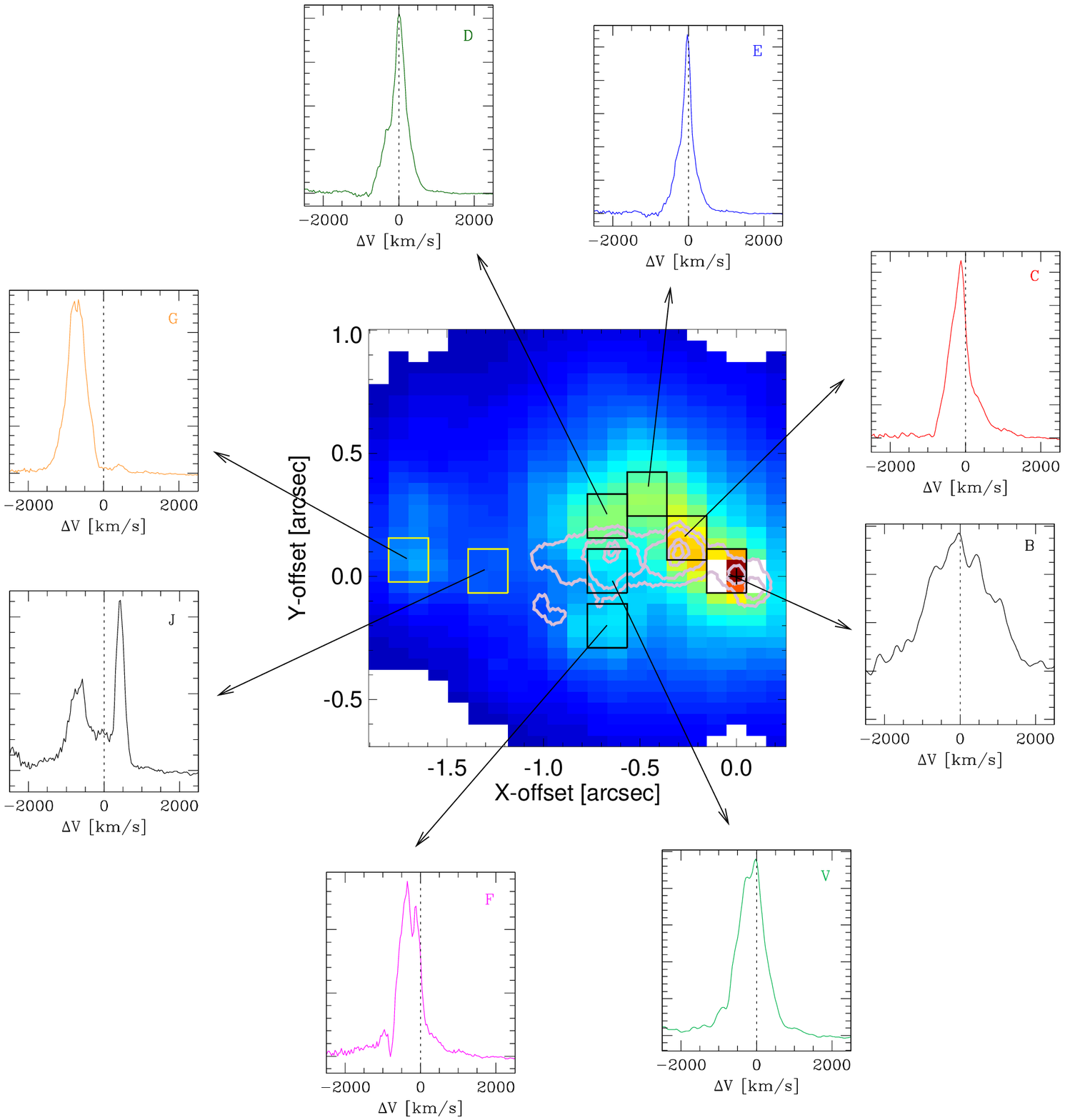}
\caption {\SiVI\ emission-line profiles. The spectra were extracted from different spatial regions of $\sim 0.2 \times 0.2~{\rm arcsec}^2$ as indicated in the figure.}
\label{f_perfiles}
\end{figure*}

\begin{figure*}
\centering
\includegraphics[width=12cm]{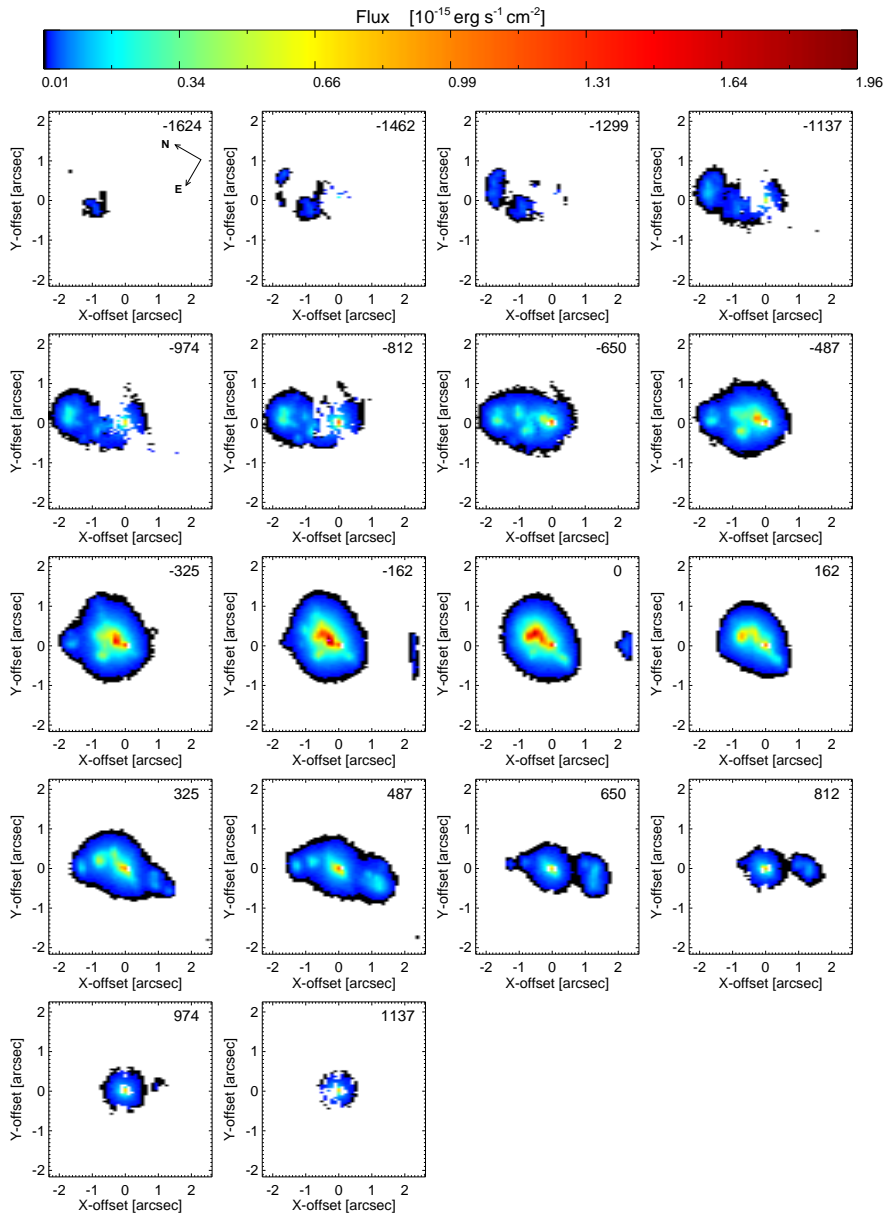}
\caption {\SiVI\ channel maps. The centre of each velocity bin relative to the systemic velocity of the galaxy is indicated in the upper right corner of the corresponding panel.}
\label{f_Si6chmaps}
\end{figure*}

\begin{figure*}
\centering
\includegraphics[width=12cm]{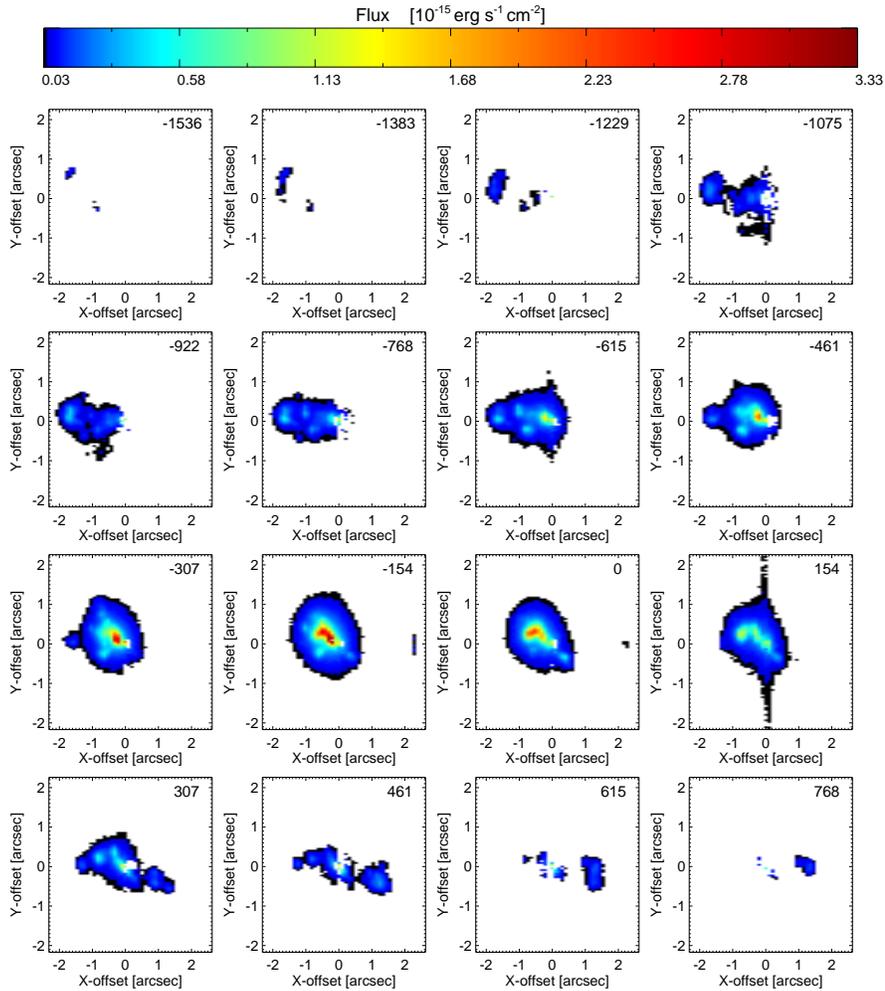}
\caption {Same as Fig.\ref{f_Si6chmaps}, but for \SiVII. Weak contamination by \H2\ emission can be seen in the panels corresponding to $\Delta {\rm v} = -1075$ and $-922$\kms\ at $\sim 1$~arcsec E from the nucleus. Additionally, bad sky subtraction affects the channels maps in regions crossing the nucleus along the y-axis, especially noticeable at $\Delta {\rm v} = 154$\kms.}
\label{f_Si7chmaps}
\end{figure*}

\begin{figure*}
\centering
\includegraphics[width=12cm]{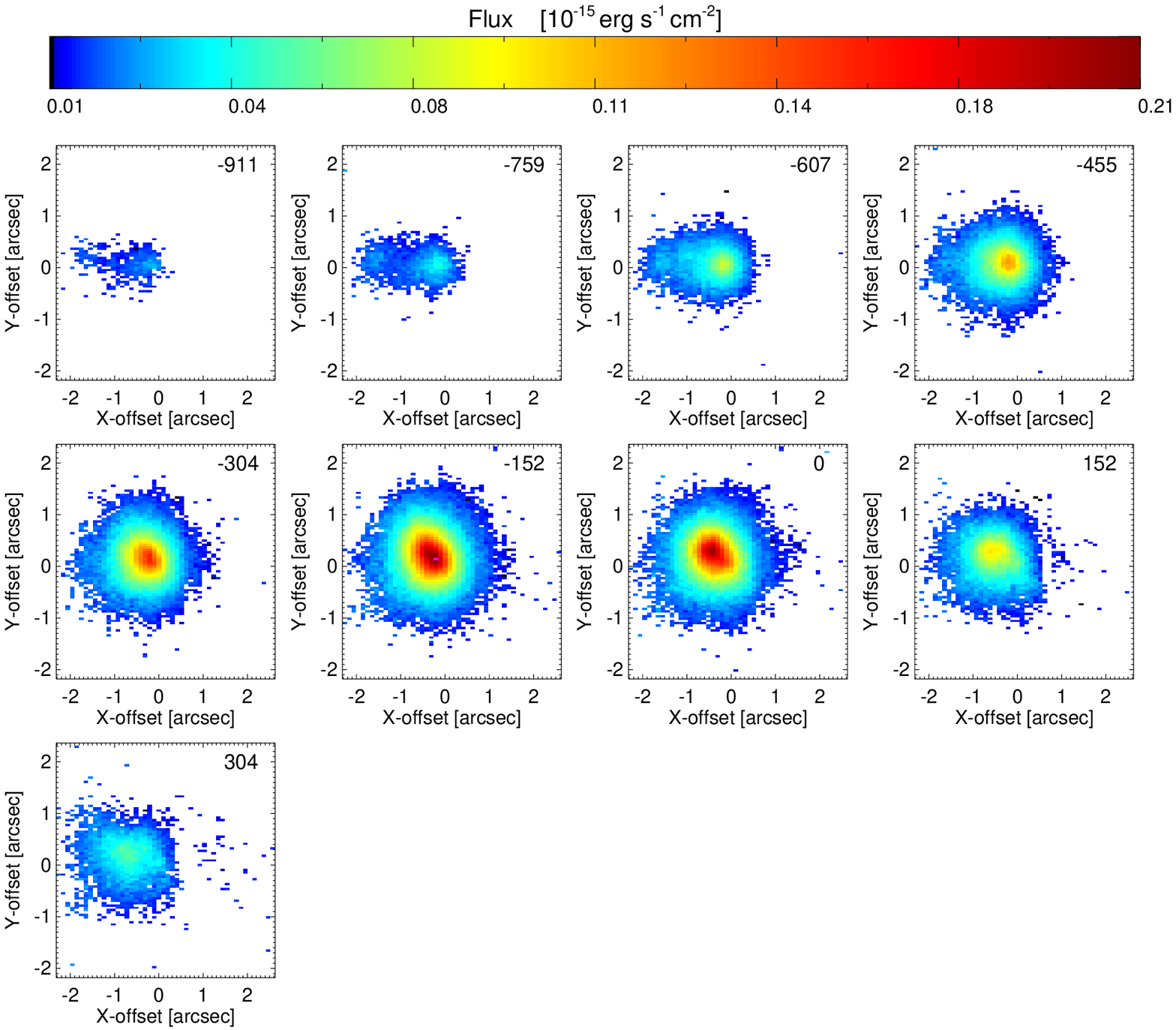}
\caption {Same as Fig.\ref{f_Si6chmaps}, but for \SIX.}
\label{f_S9chmaps}
\end{figure*}

The CLs of \n1068 display very intricate profiles, with multiple components and strong variations in their form and width from region to region. As an example, Fig.~\ref{f_perfiles} shows the \SiVI\ line profiles of spectra extracted from different spatial regions of $\sim 0.2 \times 0.2~{\rm arcsec}^2$. Regions B--G were selected to approximately match the individual clouds identified by \citet{Evans91} in the \OIII\ image (see panel e of Fig.~\ref{f_mapascontornos}).

In order to study the kinematics of the high-ionization gas surrounding the nucleus of \n1068, we have constructed channel maps of the strongest CLs (i.e. \SIX, \SiVI\ and \SiVII) observed in the NIR spectra. These maps were constructed by integrating the flux in small velocity bins ($\sim 150$\kms) along the line profiles. 
Note that the high complexity displayed by the emission lines in this galaxy does not allow any meaningful kinematic analysis in terms of the standard velocity and velocity dispersion maps derived from the Gaussian fitting of the lines.

Figs.~\ref{f_Si6chmaps} and \ref{f_Si7chmaps} show the channel maps of \SiVI\ and \SiVII, respectively. The central velocity of the integrated bin with respect to the systemic velocity of the galaxy is indicated in the upper right corner of each panel. The positive/negative values correspond to redshifted/blueshifted velocities.
It can be seen that the channel maps of both lines are very similar. Some of the observed differences can be attributed to the contamination of the \SiVII\ line by sky line residuals and \H2~2.48\micron\ emission (see section~\ref{s_nirspectrum}). Bad sky subtraction is noticeable in the channels with $\Dv=154, -615$ and $-1075$\kms, crossing the nucleus along the SE-NW direction. Contamination by \H2\ is clearly distinguishable in the panels corresponding to $\Dv = -1075$ and $-922$\kms, in the region located at $\sim 1$~arcsec E from the nucleus, where the \H2\ emission reaches its maximum (see Fig.~\ref{f_mapasLIL}). Despite these differences, the agreement in the distributions is excellent.

It is easy to see from Figs.~\ref{f_Si6chmaps} and \ref{f_Si7chmaps} that both \SiVI\ and \SiVII\ are emitted by gas with a wide range of velocities, both positives and negatives, reaching values as high as $\sim -1500$ and $1000$\kms. The gas in the central regions (inner $\sim 0.5$~arcsec) is more turbulent, emitting at practically all the observed velocities. The line profiles of these regions are broad, showing multiple peaks and strong wings.
The CL emitting gas located further out from the nucleus is aligned in the NE-SW (PA$\sim 30$\deg) direction, with a trend of the gas located towards the NE to show negative velocities while the gas located towards the SW displays positive velocities.
This trend becomes stronger as we look further from the nucleus, with an increase of the values of the observed velocities. Additionally, part of the gas located at the NE also shows positive velocities, so that the line profiles in these regions are split into two components, with one redshifted and the other one blueshifted (see Fig.~\ref{f_perfiles}).
This general pattern is not followed by the \SiVI\ and \SiVII\ emitting gas located at $\sim 2.2$~arcsec towards the SW from the nucleus, where the gas velocity is very close to that of the systemic velocity of the galaxy. Moreover,  this system seems to be disconnected from the parcel of gas located next to it, towards the nucleus, because the latter displays much higher velocities. Overall, the \SiVI\ channel maps are consistent with the ones presented by MS11 (see their fig.~12), with the exception of the higher blueshifted velocity channels, which are contaminated by \H2\ emission (see especially the two extremely strong blobs of emission in the $-1400$~\kms\ velocity bin).

The kinematics of the \SiVI\ and \SiVII\ emitting gas derived in this work is compatible with that derived by \citet{Mazzalay10} for the optical CLs along PA=22\deg. Although it is not possible to make a point-to-point comparison between the two data sets (due to astrometric imprecisions), we conclude that both optical and NIR CLs follow a similar behaviour: the nuclear lines show broad profiles with velocities close to the systemic velocity; further out, towards the SW, the strongest emission is produced by gas with receding velocities. Towards the NE, the line profiles are split into at least two components, with both receding and approaching velocities.

Fig.~\ref{f_S9chmaps} shows the channel map corresponding to the \SIX\ emission. It can be seen that this CL follows a similar trend as those of \SiVI\ and \SiVII. Note, however, that in the former the emission appears more blurred than in the latter two due to a loss of spatial resolution in the $J$-band (see Section~\ref{s_results}). Furthermore, the \SIX\ emission displays a smaller range of velocities than the silicon lines. The high-velocity gas (both approaching and receding components) in the silicon lines is mostly found in regions far from the nucleus (both towards the NE and SW), where weak or no \SIX\ emission is observed.

Overall, the integrated flux map of the CLs presents a clear asymmetry along the SW-NE direction (PA~$\sim 30$\deg), that is, aligned with the radio emission observed with VLA (see panel a of  Fig.~\ref{f_mapascontornos}). However, when we look at the channel velocity maps, it is possible to discriminate two different behaviours: the high-velocity gas is extended along the SW-NE direction while the low-velocity gas ($\Dv\sim -$300--100~\kms) displays an asymmetry along the S-N direction. The latter corresponds to the original direction of the jet (formed by the radio knots S1 and S2) before being deflected to its final direction (PA$\sim 30$\deg).
Assuming that the jet is initially aligned to the collimation axis of the ionizing radiation (i.e. the S-N direction) before bending to the NE, we expect that photoionization by the central source will be more relevant in that direction. In contrast, processes related to the interaction of the jet and its surrounding medium should be enhanced along the NE-SW direction. Taking this into account, it is possible to associate the low-velocity CL emission to gas photoionized by the AGN while gas radially accelerated due to the interaction with the radio jet should display the largest velocities. Indeed, Figs.~\ref{f_Si6chmaps} and~\ref{f_Si7chmaps} show that gas accelerated to velocities of up to 500~\kms\ is present along the edges of the radio contours. This scenario would also account for the blue- and redshifted lines observed to the NE and SW of the nucleus, respectively.

Currently, two main competing models have been proposed to explain the kinematics of the ionized gas in the nucleus of \n1068: on the one hand, the observed NLR kinematics are the result of the interaction of the nuclear gas and the radio jet \citep[e.g.][]{Axon97, Capetti97, Axon98}. On the other hand, the observed kinematics can be explained by a biconical radial outflow, pointing to scenarios where the gas acceleration is produce by radiation and/or wind pressure \citep[e.g.][]{Crenshaw00a, Das06}.
While these models suggest very different scenarios, the observational data obtained to date are consistent with both models.
Very recently, MS11 tested the scenario proposed by \citet{Crenshaw00a} using \SiVI\ channel maps derived from integral field spectroscopy SINFONI/VLT observations. Although the detailed description of the kinematic modelling method was left for a future publication, these authors find that the best-fitting  model parameters are in agreement with those of previous long-slit HST studies \citep{Crenshaw00a, Das06}.
Note, however, that the derived parameters may be affected by the contamination of the \H2~1.96\micron\ to the \SiVI\ emission line discussed above, which particularly affects the blueshifted high-velocity channels.

The overall trend observed in the CLs channel maps derived from the NIFS observations is consistent with the main kinematic models proposed so far to explain the long-slit observations of the ionized gas in this galaxy. However, the richness of the 3D data presented here gives a more global view of the nucleus of \n1068, and will guide future modelling.

\subsection{Emission-line ratios}\label{s_lineratios}

Emission-line ratios provide important information about physical properties of the emission-line gas (e.g. density, temperature and metallicity) averaged over all column densities along a particular line of sight.
The excellent spatial resolution and broad wavelength coverage of the integral field spectroscopy data allow us to alleviate the limitations caused by projection effects by the analysis of the emission-line ratios produced by gas with different velocities. Although it is not possible to guarantee that parcels of gas with similar velocity are co-spatial, we can assert that parcels of gas with very different velocities are located in different regions along the line of sight. Thus, the resultant physical parameters are more representative of the different regions than the ones we would obtain integrating along the line of sight. In this section, we analyse the \SiVII/\SiVI, \FeII/\PaB\ and \FeII/\PII\ emission-line ratios in order to gain information about the ionization structure and the presence of shocks in the inner regions of \n1068.

\subsubsection{The \SiVII/\SiVI\ line ratio}\label{s_si7_Si6}

\begin{figure*}
\centering
\includegraphics[width=12cm]{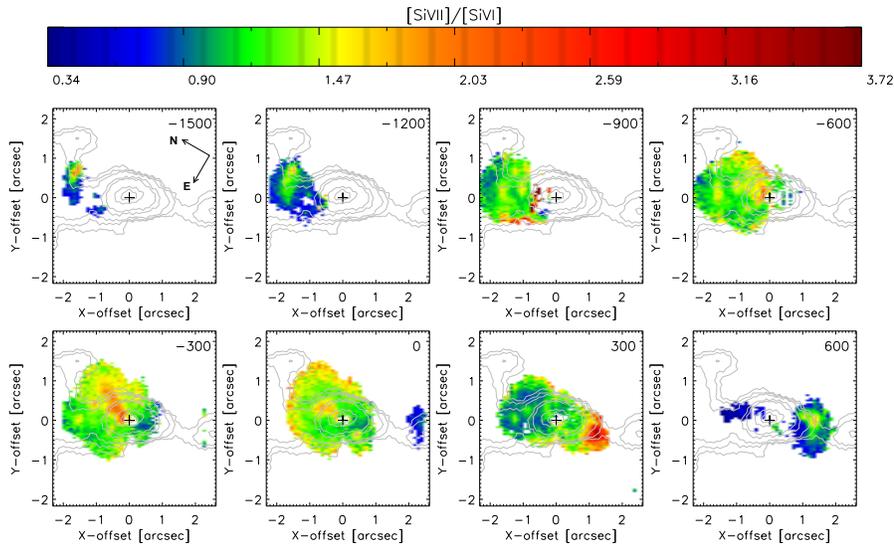}
\caption {\SiVII/\SiVI\ line ratio distribution measured for different velocity bins. The central velocity of each bin (relative to the systemic velocity) is indicated in the upper right corner of the corresponding panel. }
\label{f_Si7Si6}
\end{figure*}

Few works in the literature have reported the simultaneous detection of \SiVI\ and \SiVII\ in \n1068. To our knowledge, only \citet{Marconi96} and \citet{Rodriguez-Ardila06b} measured these two lines by means of long-slit spectroscopy. The relative strengths of the silicon emission lines found in these two works pointed to a hard spectrum photoionizing source as the primary excitation for the majority of the emission. It was assumed that the source was the nuclear central engine. Moreover, \citet{Rodriguez-Ardila06b} suggest that shocks generated by the radio jet NLR interaction could provide the additional power required for line formation. Individual \SiVI\ and \SiVII\ images, shown in Figs.~\ref{f_Si6chmaps} and \ref{f_Si7chmaps}, respectively, reveal resolved and extended emission in both lines, with at least three major separate components: one close to the nucleus, following the direction of the radio jet; a second one at spot `G' of \citet{Evans91} and a third at 2~arcsec SW from the nucleus. Note that the latter two appear significantly stronger than or comparable to \BrG\ emission.

As both silicon lines are from the same element, their ratio is independent of the metal abundance. Because of their difference in ionization potential ($\Delta{\rm IP}\sim40$~eV) the ratio can be used to map the ionization structure and mechanisms powering the CLs in greater detail than previous long-slit spectroscopy.

Fig.~\ref{f_Si7Si6} shows the \SiVII/\SiVI\ line flux ratio in \n1068 for the velocity bins between -1500\kms\ to 600\kms\ at a step of 300\kms. Only ratio values with S/N$>$3 are plotted. It can be seen that a large variation in the value of the ratio is found across the different regions, from $\sim 0.4$ to $\sim 3$, that is, nearly 1 dex. The largest values are found in an elongated region coincident with the inner portion of the jet before bending (S-N direction), where a ratio of $\sim 1.9$ is measured, and at about 1.3~arcsec SW from the centre \citep[cloud HST-H;][]{Wang12}, where a bright compact spot displaying a ratio of $\sim 3$ is observed. In velocity space, these two peaks correspond to gas at moderate blueshift (${\rm v}=-300$\kms) and redshift (${\rm v}=300$\kms), respectively. The `G' spot at $\sim 130$~pc NE from the nucleus appears as an isolated blueshifted feature with a \SiVII/\SiVI\ line flux ratio peaking at $\sim 1.5$. The smaller values (\SiVII/\SiVI\ $\sim 0.5$) are predominantly found around the `G' spot as well as in the emission region at 2.2~arcsec SW from the centre, and in some localized regions SW of the nucleus. In most other areas, the line ratio varies between 0.9 and 1.4. For reference, \citet{Marconi96} and \citet{Rodriguez-Ardila06b} measured a \SiVII/\SiVI\ integrated flux line ratio of $0.8\pm0.3$ and $0.6\pm0.2$, respectively.

Taking into account uncertainties, the flux ratios measured in the GEMINI/NIFS data agree with the ones measured previously in the long-slit data except for the three peaks mentioned above. The fact that they do not coincide spatially with the nucleus, or are not symmetrically distributed around it,  reflects the complexity of the inner regions of \n1068, where, for example, an anisotropic radiation field, gradients in density and/or selective absorption could be shaping the observed ionization structure.

\citet{Ferguson97b} present the results of a large number of photoionization simulations of the CLR in AGNs, with  the ionization parameter $U(H)$\footnote{The ratio of ionizing photon density to Hydrogen density, $U(H) = \Phi(H)/n(H)c$, where $\Phi (H)$ is the flux of ionizing photons and $c$ is the speed of light.} being the fundamental parameter of their models. They found that CLs form at distances from just outside the broad-line region to $\sim400L_{43.5}^{1/2}$~pc, where $L_{43.5}$ is the ionizing luminosity in units of $10^{43.5}$~ergs~s$^{-1}$, in gas with ionization parameter $-2.0\lesssim {\rm log~} U(H) \lesssim 0.75$, suggesting that CLs will form close to the nucleus in high-density gas and further out from the nucleus in lower density gas. The peak equivalent width of each line is provided by the models. Since that quantity is referenced to the same point in the incident continuum, a comparison should indicate grossly the expected relative strengths of the CLs. Using table~1 of \citet{Ferguson97b}, a \SiVII/\SiVI\ ratio of 1.2 is predicted. At first sight, this result agrees, within a factor of 2, with the values observed in \n1068. Note, however, that the calculation refers to the peak in the equivalent width distribution, which occurs at distances of a few parsecs from the central source and gas density $>10^6$~cm$^{-3}$.  At distances of about 100~pc, the ratio is expected to drop to a fraction of the peak value.

The above result reinforces the need of an additional mechanism powering the CL emission in \n1068. In the light of all observational evidence seen in the preceding sections, we are therefore let to consider the role of shocks, produced by interactions of either the radio jet or a radially accelerated outflow and the ISM. Fig.~\ref{f_Si7Si6} shows, for instance, that spot `G' is particularly enhanced in the blueshifted bins (v$=-900$ and $-600$\kms), with a \SiVII/\SiVI\ ratio of $\sim 1.5$ at the centre of the spot, surrounded by gas characterized by a line ratio $<1$. As the NE region of \n1068 is very little affected by extinction \citep[e.g.][]{Kraemer00c, Kraemer00b}, dust cannot be blamed for the increase of \SiVII\ emission compared to that of \SiVI. Moreover, spot `G' is located at $\sim 130$~pc from the centre, which according to the results of \citet{Ferguson97b} is far from the outer radius where \SiVII\ emission is expected to occur ($< 80$~pc) assuming photoionization by the central source alone. Cloud `HST-H' is also peculiar in this respect, as it shows the largest \SiVII/\SiVI\ ratio ($\sim 3$) and is located at a similar distance as spot `G' but in the opposite direction from the nucleus.
Wang et al. (2012) found that both spot `G' and cloud `HST-H' also share similar properties at higher energies: both are conspicuous in X-ray emission and show significantly lower [O\,{\sc iii}]/X-ray ratio ($\sim 1$) than the typical range spanned by photoionized clouds ($\sim$3-11). They interpreted this as a higher collisional ionization at these locations and suggest that these clouds are strongly interacting with the jet, and that shocks driven into the obstructing clouds produce thermal X-ray emission. The increase in the \SiVII/\SiVI\ ratio at these two locations reinforces this hypothesis.
 
In this regard, it is very useful to check if models involving shocks are able to reproduce the observed \SiVII/\SiVI\ ratios of \n1068. \citet[CV01 hereafter]{Contini01}, for instance, presented a grid of simulations where the effect of shocks coupled to photoionization by the central source is employed to calculate the spectrum emitted by the NLR of AGNs. In this approach, the clouds are moving outwards from the galaxy centre, with the shock front forming on the outer edge of the cloud, whereas the ionizing radiation reaches the opposite edge that faces the active centre. The ionization due to the primary radiation (from the central source), to the diffuse radiation generated by the free-free and free-bound transitions of the shocked and photoionized gas, as well as the collisional ionization, are all accounted for. The shock velocity $V_{\rm s}$ and the ionizing flux from the central source at the Lyman limit reaching the cloud, $F_{\rm h}$ (in units of cm$^{-2}$~s$^{-1}$~eV$^{-1}$), are the main input parameters. Other parameters include the pre-shock density, $n_0$, and the pre-shock magnetic field, $B_0$.

Tables~1--12 of CV01 show that shock-dominated clouds ($F_{\rm h}=0$) with shock velocities in the range $300-500$\kms\ strongly favour the production of \SiVII/\SiVI\ ratios larger than 1.5 (models 56 and 70 for example). Moreover, model 70, with $V_{\rm s}=500$\kms, produces a line flux ratio of $\sim 3$. For all these clouds, a solar metallicity, $n_0 =300$~cm$^{-3}$ and $B_0 = 10^{-4}$ Gauss were adopted. When coupled to the presence of a strong radiation field from the central source (e.g. ${\rm log~}F_{\rm h} =12$, model 62), the value of the ratio drops to $\sim 0.8$. Thus, shocks locally enhance the emission of CLs even when taking into account its photoionized precursor. As the maps of Fig.~\ref{f_Si7Si6} show, blueshifted emission lines in the interval $300 -900$\kms\ characterize the regions where the largest \SiVII/\SiVI\ ratios are observed. That is, well within the range of shock velocities considered in the CV01 models. Note also that the gas is compressed downstream depending on $V_{\rm s}$. For a $V_{\rm s} =300$\kms\ and $n_0 =300$~cm$^{-3}$, the maximum downstream density  $n_{\rm max}$ is $1.04\times10^{4}$~cm$^{-3}$ while for $V_{\rm s} =500$\kms, $n_{\rm max}$ is $1.75\times10^{4}$~cm$^{-3}$ (see table 13 of CV01). In both cases, $n_{\rm max}$ is consistent with the upper limit for the CLR found in Section~\ref{s_morphology}.

Additional support of the contribution of shocks to the CL emission in \n1068 can be obtained from the work of \citet{Allen08}. They published an extensive library of fully radiative shock models that include both the radiative shock and its photoionized precursor. Similar to the models of CV01, the shock velocity, the pre-shock density and the intensity of the magnetic field are the main input parameters. Model predictions for the composite shock$+$precursor structure for the \SiVII/\SiVI\ ratio are found to vary from 0.5 to 1.2 for velocities between 500 and 1000\kms, respectively, and $n_0$ of 100~cm$^{-3}$. Using this approach, \citet{Exposito11} were able to reproduce the CLs at the different nodules in \n1068, which they attributed to a local jet-induced ionizing continuum. We should keep in mind, though, that the shock model predictions provide us with only a first-order approximation. In order to have a full description of the CLR based on model-fitting, we should suitably combine clouds under different physical conditions (i.e., clouds with different $V_{\rm s}$ and $n_0$) with the proper weight so that the emission lines and the observed continuum can be reproduced. This is far beyond the scope of this paper and is left for a future publication. What is most important here is that shock models are able to reproduce CL flux ratios at distances as far as a few hundred parsecs, where photoionization by the central source predicts no or very faint CL emission.

It is also very interesting to comment on the high-excitation channel with origin at the central engine and extending to the north, displaying a nearly constant \SiVII/\SiVI\ ratio of $\sim 1.9$ (see Fig.~\ref{f_Si7Si6} at the ${\rm v}=-300$\kms\ bin). This feature is coincident with the position angle of the radio jet before bending, continuing with that orientation beyond the point where the radio jet changes its orientation. At the distances from the centre at which this gas is located ($\lesssim 1$~arcsec), photoionization by the central source should be enough to power the observed CLs. Note, however, that faint X-ray emission is also seen in the images of \citet{Wang12} along the high-excitation channel, revealing the presence of hot plasma. Shocks along this region should not be discarded.

All the observational evidence above points out that shocks must contribute significantly to CL emission,  confirming early suggestions based on long-slit observations \citep[e.g.][]{Rodriguez-Ardila06b}. Previous works had considered this possibility but overall found that photoionization by the central source is the main driving mechanism of the CL emission \citep[e.g.][]{Oliva90, Marconi96, Kraemer00b, Mazzalay10}. The evidence shown here observationally confirms the relevance of shocks as an important component of the CL emission in \n1068 and, thanks to the spatial and spectral coverage of our integral field spectroscopic data, we are able to identify the particular regions where shocks play a significant role.
A comparison with the exquisite X-ray map of the inner 200~pc of that galaxy is in accord with this scenario.

In summary, the \SiVII/\SiVI\ line flux ratio map found for \n1068 reveals localized regions where the gas excitation cannot be easily explained by simple photoionization by the central source. Moreover, the fact that these regions follow closely the radio morphology at some specific locations, invokes a scenario where shocks originating from the interaction of the emitting clouds with the radio jet contribute to enhance the ionization state of the gas. However, other sources of shocks should also be considered (the interaction of outflowing clouds with the ISM, for instance). Models taking into account the effect of shocks, with velocities similar to those measured in \n1068, can account for the high \SiVII/\SiVI\ line flux ratio observed at some locations in the NLR of that object.

\subsubsection{The role of shocks in the formation of \FeII: the \FeII/\PaB\ and \FeII/\PII\ line ratios}\label{s_feii}

\begin{figure*}
\centering
\includegraphics[width=12cm]{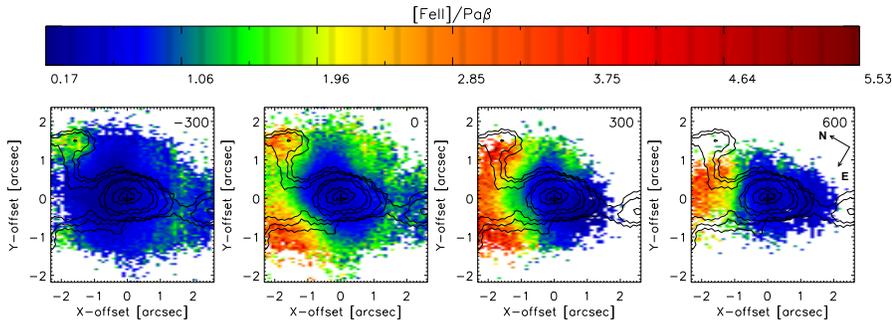}
\caption {\FeII/\PaB\ line ratio distribution measured for different velocities bins. The central velocity of each bin (relative to the systemic velocity) is indicated in the upper right corner of the corresponding panel.}
\label{f_Fe2PaB}
\end{figure*}

\begin{figure*}
\centering
\includegraphics[width=12cm]{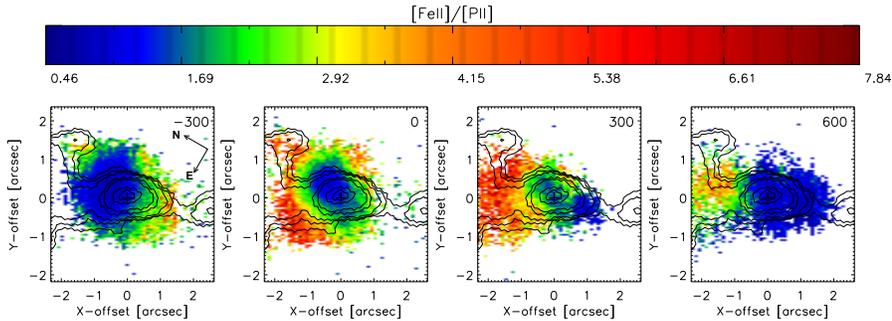}
\caption {\FeII/\PII\ line ratio distribution measured for different velocities bins. The central velocity of each bin (relative to the systemic velocity) is indicated in the upper right corner of the corresponding panel.}
\label{f_Fe2P2}
\end{figure*}

The origin of the \FeII\ emission in galaxies has been extensively studied \citep[e.g.][]{Forbes93, Simpson98, Rodriguez-Ardila04, Rodriguez-Ardila05b, Storchi-Bergmann09}. However, the excitation mechanisms and location of the \FeII\ emitting gas are still a subject of debate. Due to the 
low-ionization potential (16.2~eV), the general consensus is that \FeII\ is formed in a partially ionized zone. Photoionization by X-rays and shocks have been suggested as the most important mechanisms contributing to the formation of the \FeII\ in this region.

In this context, the \FeII/\PaB\ ratio is considered as an important tracer of the origin of the \FeII\ emission. There is an increasing progression in this ratio from pure photoionization to pure shock excitation \citep{Simpson98} as one goes from starbursts (displaying typical values of \FeII/\PaB\ $\sim 0.5$) through AGNs to SNRs and shocks (with values higher than $\sim 2$). Therefore, this line ratio can be used to measure the relative contribution of photoionization and shocks in the formation of the \FeII\ lines. Fig.~\ref{f_Fe2PaB} shows the \FeII/\PaB\ line ratio calculated for four different velocity ranges in the nucleus of \n1068, with the 6~cm VLA contours overlaid. Low values of \FeII/\PaB\ ($< 2$) and a homogeneous distribution characterize the gas with velocities in the range $\Dv = -300 \pm 150$~\kms. On the other hand, the gas with velocities close to the systemic velocity and higher show a very different line ratio morphology, resembling partially concentric rings with a symmetry axis coincident with the radio jet. The values of \FeII/Pa$\beta$ increase in regions further out from the nucleus, reaching values in the SNRs range ($\gtrsim 2$) in the regions associated with the NE end of the jet. Both the shape and values of the \FeII/\PaB\ strongly suggest the presence of shocks in the inner regions of \n1068. These contribute to the formation of \FeII\ in the outer regions while photoionization probably dominates closer to the nucleus.

Another useful tool to constrain the origin of \FeII\ lines is the \FeII~1.26~\micron/\PII~1.19~\micron\ line ratio. \cite{Oliva01} proposed that in objects with low \FeII/\PII\ ratios ($\lesssim 2$), shocks do not play an important role in the line excitation while large values of this ratio ($\gtrsim 20$) indicate that the emitting gas has recently passed through a fast shock and it is likely that the lines are produced by shock excitation. Although the \PII\ emission in \n1068 is weak, it was possible to construct \FeII/\PII\ maps for the innermost regions. These maps are presented in Fig.~\ref{f_Fe2P2}. The measured values of this line ratio vary between $\sim 0.5 - 7.8$ and, as in the case of the \FeII/\PaB\ ratio, its distribution shows a spatial  correlation with the radio jet. Although the measured values are lower than the value of 20 derived by \cite{Oliva01} for shocked gas, the regions associated with the NE component of the jet are much higher than the lower limit of 2 suggested by these authors, indicating an important increase in the relative abundance of Fe/P due to shocks.

Additional evidence for both photoionization and radio jet/shock excitation of \FeII\ in the nuclei of AGNs can be found, for example, in the work presented by \citet{Storchi-Bergmann09, Storchi-Bergmann10}. These authors analysed NIFS/Gemini integral field spectroscopy data of the well-known Seyfert~1 galaxy \n4151, finding a strong influence of the radio jet present in the centre of this galaxy on the \FeII\ emission. From a similar analysis to the one presented here, these authors concluded that photoionization dominates in the inner region while shocks dominate in the outer region of the galaxy.

A more detailed discussion on this subject, together with the analysis of the low-ionization lines of \n1068, is carried out by Riffel et al. (in preparation) using the same set of data.

In summary, our findings indicate that shocks produced by the interaction between the radio jet and the ambient gas are responsible for most of the \FeII\ emission observed in the nucleus of \n1068, especially in the regions located at the NE of the nucleus.

\section{Summary and conclusions}\label{s_summary}

In this paper we have presented new $J$- and $K$-band 2D spectra of the inner $\sim 300$~pc of \n1068. The data were  obtained with the Gemini NIFS integral field unit spectrometer, which provided us with high spatial and spectral resolution sampling.

Our main results can be summarized as follows.
\begin{enumerate}
      \item The CLs observed in the spectra of \n1068, \SiVI, \SiVII, \SIX, \CaVIII\ and \AlIX, display very asymmetric and inhomogeneous distributions. The emission is elongated along the NE-SW direction, with the stronger emission preferentially localized to the NE of the nucleus. We observe CL emission up to a distance of $\sim 170$~pc to the NE and SW of the nucleus.

      \item Hydrogen and \OIII\ low-ionization lines share a similar distribution (bright knots and arcs) as the CLs, suggesting a common origin. On the other hand, the morphology displayed by the molecular hydrogen lines is completely different. The \FeII\ emission appears to have a shock-excited component.

      \item CLs are emitted by gas covering a wide range of velocities, up to $\sim 1000$\kms\ relative to the systemic velocity of the galaxy. There is a trend for the gas located on the NE side of the nucleus to be  blueshifted while the gas located towards the SW is redshifted. Nevertheless, some redshifted emission is observed NE from the nucleus, consistent with the double-peak profiles observed in optical high- and low-ionization lines.

      \item The comparison of the line flux distributions with the radio emission suggests that the {\it morphology} of the emission-line gas is closely related to the radio jet observed in the nucleus of \n1068. On the other hand, although the overall gas {\it kinematics} is consistent with being driven by the radio jet, other mechanisms are also possible.

      \item The ionization structure, mapped by the \SiVII/\SiVI\ emission-line ratio, is highly intricate. While this ionization map is not easily explained by photoionization by a central source, models taking into account the effects of shocks are able to reproduce the high values of \SiVII/\SiVI\ observed in some regions.

      \item The presence of shocks in the nucleus of \n1068 is supported by the high values and distribution found for the \FeII/\PaB\ and \FeII/\PII\ line ratios.

\end{enumerate}

The richness and complexity of the structures we observe, both in terms of velocity fields and emission-line ratios are remarkable. Therefore, the use of integral field spectroscopy is essential in order to map the circumnuclear environment of nearby AGN, to disentangle the effects of photoionization and shocks on emission-line properties, and to study the effects of radio jets on their gaseous environment.

\section*{Acknowledgments}
ARA thanks to Conselho Nacional de Desenvolvimento Cient\'{\i}fico e Tecnol\'ogico, CNPq, for financial support through grant 308877/2009-8. PJM acknowledges the support of the Australian Research Council (ARC) through Discovery Project DP0342844.


\begin{thebibliography}{61}
\expandafter\ifx\csname natexlab\endcsname\relax\def\natexlab#1{#1}\fi

\bibitem[{{Allen} {et~al}\mbox{.}(2008){Allen}, {Groves}, {Dopita},
  {Sutherland}, \& {Kewley}}]{Allen08}
{Allen} M.~G., {Groves} B.~A., {Dopita} M.~A., {Sutherland} R.~S., {Kewley}
  L.~J., 2008, \apjs, 178, 20

\bibitem[{{Axon} {et~al}\mbox{.}(1998){Axon}, {Marconi}, {Capetti}, {Maccetto},
  {Schreier}, \& {Robinson}}]{Axon98}
{Axon} D.~J., {Marconi} A., {Capetti} A., {Maccetto} F.~D., {Schreier} E.,
  {Robinson} A., 1998, \apjl, 496, L75

\bibitem[{{Axon} {et~al}\mbox{.}(1997){Axon}, {Marconi}, {Macchetto},
  {Capetti}, \& {Robinson}}]{Axon97}
{Axon} D.~J., {Marconi} A., {Macchetto} F.~D., {Capetti} A., {Robinson} A.,
  1997, \apss, 248, 69

\bibitem[{{Bicknell} {et~al}\mbox{.}(1998){Bicknell}, {Dopita}, {Tsvetanov}, \&
  {Sutherland}}]{Bicknell98}
{Bicknell} G.~V., {Dopita} M.~A., {Tsvetanov} Z.~I., {Sutherland} R.~S., 1998,
  \apj, 495, 680

\bibitem[{{Binette} {et~al}\mbox{.}(1997){Binette}, {Wilson}, {Raga}, \&
  {Storchi-Bergmann}}]{Binette97}
{Binette} L., {Wilson} A.~S., {Raga} A., {Storchi-Bergmann} T., 1997, \aap,
  327, 909

\bibitem[{{Capetti} {et~al}\mbox{.}(1997){Capetti}, {Axon}, \&
  {Macchetto}}]{Capetti97}
{Capetti} A., {Axon} D.~J., {Macchetto} F.~D., 1997, \apj, 487, 560

\bibitem[{{Cecil} {et~al}\mbox{.}(2002){Cecil}, {Dopita}, {Groves}, {Wilson},
  {Ferruit}, {P{\'e}contal}, \& {Binette}}]{Cecil02}
{Cecil} G., {Dopita} M.~A., {Groves} B., {Wilson} A.~S., {Ferruit} P.,
  {P{\'e}contal} E., {Binette} L., 2002, \apj, 568, 627

\bibitem[{{Contini} {et~al}\mbox{.}(2003){Contini}, {Rodr{\'i}guez-Ardila}, \&
  {Viegas}}]{Contini03}
{Contini} M., {Rodr{\'i}guez-Ardila} A., {Viegas} S.~M., 2003, \aap, 408, 101

\bibitem[{{Contini} \& {Viegas}(2001)}]{Contini01}
{Contini} M., {Viegas} S.~M., 2001, \apjs, 132, 211

\bibitem[{{Crenshaw} \& {Kraemer}(2000)}]{Crenshaw00a}
{Crenshaw} D.~M., {Kraemer} S.~B., 2000, \apjl, 532, L101

\bibitem[{{Das} {et~al}\mbox{.}(2006){Das}, {Crenshaw}, {Kraemer}, \&
  {Deo}}]{Das06}
{Das} V., {Crenshaw} D.~M., {Kraemer} S.~B., {Deo} R.~P., 2006, \aj, 132, 620

\bibitem[{{Deo} {et~al}\mbox{.}(2007){Deo}, {Crenshaw}, {Kraemer}, {Dietrich},
  {Elitzur}, {Teplitz}, \& {Turner}}]{Deo07}
{Deo} R.~P., {Crenshaw} D.~M., {Kraemer} S.~B., {Dietrich} M., {Elitzur} M.,
  {Teplitz} H., {Turner} T.~J., 2007, \apj, 671, 124

\bibitem[{{Dopita} \& {Sutherland}(1996)}]{Dopita96}
{Dopita} M.~A., {Sutherland} R.~S., 1996, \apjs, 102, 161

\bibitem[{{Dressel} {et~al}\mbox{.}(1997){Dressel}, {Tsvetanov}, {Kriss}, \&
  {Ford}}]{Dressel97}
{Dressel} L.~L., {Tsvetanov} Z.~I., {Kriss} G.~A., {Ford} H.~C., 1997, \apss,
  248, 85

\bibitem[{{Evans} {et~al}\mbox{.}(1991){Evans}, {Ford}, {Kinney}, {Antonucci},
  {Armus}, \& {Caganoff}}]{Evans91}
{Evans} I.~N., {Ford} H.~C., {Kinney} A.~L., {Antonucci} R.~R.~J., {Armus} L.,
  {Caganoff} S., 1991, \apjl, 369, L27

\bibitem[{{Exposito} {et~al}\mbox{.}(2011){Exposito}, {Gratadour},
  {Cl{\'e}net}, \& {Rouan}}]{Exposito11}
{Exposito} J., {Gratadour} D., {Cl{\'e}net} Y., {Rouan} D., 2011, \aap, 533,
  A63

\bibitem[{{Fabian}(2012)}]{Fabian12}
{Fabian} A.~C., 2012, \araa, 50, 455

\bibitem[{{Ferguson} {et~al}\mbox{.}(1997){Ferguson}, {Korista}, \&
  {Ferland}}]{Ferguson97b}
{Ferguson} J.~W., {Korista} K.~T., {Ferland} G.~J., 1997, \apjs, 110, 287

\bibitem[{{Forbes} \& {Ward}(1993)}]{Forbes93}
{Forbes} D.~A., {Ward} M.~J., 1993, \apj, 416, 150

\bibitem[{{Galliano} \& {Alloin}(2002)}]{Galliano02}
{Galliano} E., {Alloin} D., 2002, \aap, 393, 43

\bibitem[{{Galliano} {et~al}\mbox{.}(2003){Galliano}, {Alloin}, {Granato}, \&
  {Villar-Mart{\'i}n}}]{Galliano03}
{Galliano} E., {Alloin} D., {Granato} G.~L., {Villar-Mart{\'i}n} M., 2003,
  \aap, 412, 615

\bibitem[{{Gallimore} {et~al}\mbox{.}(1996){Gallimore}, {Baum}, {O'Dea}, \&
  {Pedlar}}]{Gallimore96}
{Gallimore} J.~F., {Baum} S.~A., {O'Dea} C.~P., {Pedlar} A., 1996, \apj, 458,
  136

\bibitem[{{Gelbord} {et~al}\mbox{.}(2009){Gelbord}, {Mullaney}, \&
  {Ward}}]{Gelbord09}
{Gelbord} J.~M., {Mullaney} J.~R., {Ward} M.~J., 2009, \mnras, 397, 172

\bibitem[{{Komossa} \& {Fink}(1997{\natexlab{a}})}]{Komossa97a}
{Komossa} S., {Fink} H., 1997{\natexlab{a}}, \aap, 322, 719

\bibitem[{{Komossa} \& {Fink}(1997{\natexlab{b}})}]{Komossa97b}
{Komossa} S., {Fink} H., 1997{\natexlab{b}}, \aap, 327, 483

\bibitem[{{Komossa} \& {Schulz}(1997)}]{Komossa97c}
{Komossa} S., {Schulz} H., 1997, \aap, 323, 31

\bibitem[{{Komossa} {et~al}\mbox{.}(2008){Komossa}, {Xu}, {Zhou},
  {Storchi-Bergmann}, \& {Binette}}]{Komossa08}
{Komossa} S., {Xu} D., {Zhou} H., {Storchi-Bergmann} T., {Binette} L., 2008,
  \apj, 680, 926

\bibitem[{{Korista} \& {Ferland}(1989)}]{Korista89}
{Korista} K.~T., {Ferland} G.~J., 1989, \apj, 343, 678

\bibitem[{{Kraemer} \& {Crenshaw}(2000{\natexlab{a}})}]{Kraemer00c}
{Kraemer} S.~B., {Crenshaw} D.~M., 2000{\natexlab{a}}, \apj, 532, 256

\bibitem[{{Kraemer} \& {Crenshaw}(2000{\natexlab{b}})}]{Kraemer00b}
{Kraemer} S.~B., {Crenshaw} D.~M., 2000{\natexlab{b}}, \apj, 544, 763

\bibitem[{{Macchetto} {et~al}\mbox{.}(1994){Macchetto}, {Capetti}, {Sparks},
  {Axon}, \& {Boksenberg}}]{Macchetto94}
{Macchetto} F., {Capetti} A., {Sparks} W.~B., {Axon} D.~J., {Boksenberg} A.,
  1994, \apjl, 435, L15

\bibitem[{{Marconi} {et~al}\mbox{.}(1994){Marconi}, {Moorwood}, {Salvati}, \&
  {Oliva}}]{Marconi94}
{Marconi} A., {Moorwood} A.~F.~M., {Salvati} M., {Oliva} E., 1994, \aap, 291,
  18

\bibitem[{{Marconi} {et~al}\mbox{.}(1996){Marconi}, {van der Werf}, {Moorwood},
  \& {Oliva}}]{Marconi96}
{Marconi} A., {van der Werf} P.~P., {Moorwood} A.~F.~M., {Oliva} E., 1996,
  \aap, 315, 335

\bibitem[{{Mazzalay} {et~al}\mbox{.}(2010){Mazzalay}, {Rodr{\'i}guez-Ardila},
  \& {Komossa}}]{Mazzalay10}
{Mazzalay} X., {Rodr{\'i}guez-Ardila} A., {Komossa} S., 2010, \mnras, 405, 1315

\bibitem[{{McGregor} {et~al}\mbox{.}(2003){McGregor}, {Hart}, {Conroy},
  {Pfitzner}, {Bloxham}, {Jones}, {Downing}, {Dawson}, {Young}, {Jarnyk}, \&
  {Van Harmelen}}]{McGregor03}
{McGregor} P.~J. {et~al.}, 2003, in Society of Photo-Optical Instrumentation
  Engineers (SPIE) Conference Series, Vol. 4841, Society of Photo-Optical
  Instrumentation Engineers (SPIE) Conference Series, {Iye} M., {Moorwood}
  A.~F.~M., eds., pp. 1581--1591

\bibitem[{{Mullaney} \& {Ward}(2008)}]{Mullaney08}
{Mullaney} J.~R., {Ward} M.~J., 2008, \mnras, 385, 53

\bibitem[{{Mullaney} {et~al}\mbox{.}(2009){Mullaney}, {Ward}, {Done},
  {Ferland}, \& {Schurch}}]{Mullaney09}
{Mullaney} J.~R., {Ward} M.~J., {Done} C., {Ferland} G.~J., {Schurch} N., 2009,
  \mnras, 394, L16

\bibitem[{{M{\"u}ller S{\'a}nchez} {et~al}\mbox{.}(2009){M{\"u}ller
  S{\'a}nchez}, {Davies}, {Genzel}, {Tacconi}, {Eisenhauer}, {Hicks},
  {Friedrich}, \& {Sternberg}}]{Muller-Sanchez09}
{M{\"u}ller S{\'a}nchez} F., {Davies} R.~I., {Genzel} R., {Tacconi} L.~J.,
  {Eisenhauer} F., {Hicks} E.~K.~S., {Friedrich} S., {Sternberg} A., 2009,
  \apj, 691, 749

\bibitem[{{M{\"u}ller-S{\'a}nchez}
  {et~al}\mbox{.}(2011){M{\"u}ller-S{\'a}nchez}, {Prieto}, {Hicks},
  {Vives-Arias}, {Davies}, {Malkan}, {Tacconi}, \& {Genzel}}]{Muller-Sanchez11}
{M{\"u}ller-S{\'a}nchez} F., {Prieto} M.~A., {Hicks} E.~K.~S., {Vives-Arias}
  H., {Davies} R.~I., {Malkan} M., {Tacconi} L.~J., {Genzel} R., 2011, \apj,
  739, 69

\bibitem[{{Nagao} {et~al}\mbox{.}(2000){Nagao}, {Taniguchi}, \&
  {Murayama}}]{Nagao00}
{Nagao} T., {Taniguchi} Y., {Murayama} T., 2000, \aj, 119, 2605

\bibitem[{{Oliva} {et~al}\mbox{.}(2001){Oliva}, {Marconi}, {Maiolino}, {Testi},
  {Mannucci}, {Ghinassi}, {Licandro}, {Origlia}, {Baffa}, {Checcucci},
  {Comoretto}, {Gavryussev}, {Gennari}, {Giani}, {Hunt}, {Lisi}, {Lorenzetti},
  {Marcucci}, {Miglietta}, {Sozzi}, {Stefanini}, \& {Vitali}}]{Oliva01}
{Oliva} E. {et~al.}, 2001, \aap, 369, L5

\bibitem[{{Oliva} \& {Moorwood}(1990)}]{Oliva90}
{Oliva} E., {Moorwood} A.~F.~M., 1990, \apjl, 348, L5

\bibitem[{{P{\'e}contal} {et~al}\mbox{.}(1997){P{\'e}contal}, {Ferruit},
  {Binette}, \& {Wilson}}]{Pecontal97}
{P{\'e}contal} E., {Ferruit} P., {Binette} L., {Wilson} A.~S., 1997, \apss,
  248, 167

\bibitem[{{Penston} {et~al}\mbox{.}(1984){Penston}, {Fosbury}, {Boksenberg},
  {Ward}, \& {Wilson}}]{Penston84}
{Penston} M.~V., {Fosbury} R.~A.~E., {Boksenberg} A., {Ward} M.~J., {Wilson}
  A.~S., 1984, \mnras, 208, 347

\bibitem[{{Pier} \& {Voit}(1995)}]{Pier95}
{Pier} E.~A., {Voit} G.~M., 1995, \apj, 450, 628

\bibitem[{{Pogge}(1988)}]{Pogge88}
{Pogge} R.~W., 1988, \apj, 328, 519

\bibitem[{{Porquet} {et~al}\mbox{.}(1999){Porquet}, {Dumont}, {Collin}, \&
  {Mouchet}}]{Porquet99}
{Porquet} D., {Dumont} A.-M., {Collin} S., {Mouchet} M., 1999, \aap, 341, 58

\bibitem[{{Prieto} {et~al}\mbox{.}(2005){Prieto}, {Marco}, \&
  {Gallimore}}]{Prieto05}
{Prieto} A.~M., {Marco} O., {Gallimore} J., 2005, \mnras, 364, L28

\bibitem[{{Riffel} {et~al}\mbox{.}(2008){Riffel}, {Storchi-Bergmann}, {Winge},
  {McGregor}, {Beck}, \& {Schmitt}}]{Riffel08}
{Riffel} R.~A., {Storchi-Bergmann} T., {Winge} C., {McGregor} P.~J., {Beck} T.,
  {Schmitt} H., 2008, \mnras, 385, 1129

\bibitem[{{Rodr{\'i}guez-Ardila} {et~al}\mbox{.}(2004){Rodr{\'i}guez-Ardila},
  {Pastoriza}, {Viegas}, {Sigut}, \& {Pradhan}}]{Rodriguez-Ardila04}
{Rodr{\'i}guez-Ardila} A., {Pastoriza} M.~G., {Viegas} S., {Sigut} T.~A.~A.,
  {Pradhan} A.~K., 2004, \aap, 425, 457

\bibitem[{{Rodr{\'i}guez-Ardila} {et~al}\mbox{.}(2011){Rodr{\'i}guez-Ardila},
  {Prieto}, {Portilla}, \& {Tejeiro}}]{Rodriguez-Ardila11}
{Rodr{\'i}guez-Ardila} A., {Prieto} M.~A., {Portilla} J.~G., {Tejeiro} J.~M.,
  2011, \apj, 743, 100

\bibitem[{{Rodr{\'i}guez-Ardila} {et~al}\mbox{.}(2006){Rodr{\'i}guez-Ardila},
  {Prieto}, {Viegas}, \& {Gruenwald}}]{Rodriguez-Ardila06b}
{Rodr{\'i}guez-Ardila} A., {Prieto} M.~A., {Viegas} S., {Gruenwald} R., 2006,
  \apj, 653, 1098

\bibitem[{{Rodr{\'i}guez-Ardila} {et~al}\mbox{.}(2005){Rodr{\'i}guez-Ardila},
  {Riffel}, \& {Pastoriza}}]{Rodriguez-Ardila05b}
{Rodr{\'i}guez-Ardila} A., {Riffel} R., {Pastoriza} M.~G., 2005, \mnras, 364,
  1041

\bibitem[{{Rodr{\'i}guez-Ardila} {et~al}\mbox{.}(2002){Rodr{\'i}guez-Ardila},
  {Viegas}, {Pastoriza}, \& {Prato}}]{Rodriguez-Ardila02a}
{Rodr{\'i}guez-Ardila} A., {Viegas} S.~M., {Pastoriza} M.~G., {Prato} L., 2002,
  \apj, 579, 214

\bibitem[{{Schinnerer} {et~al}\mbox{.}(2000){Schinnerer}, {Eckart}, {Tacconi},
  {Genzel}, \& {Downes}}]{Schinnerer00}
{Schinnerer} E., {Eckart} A., {Tacconi} L.~J., {Genzel} R., {Downes} D., 2000,
  \apj, 533, 850

\bibitem[{{Simpson} \& {Meadows}(1998)}]{Simpson98}
{Simpson} C., {Meadows} V., 1998, \apjl, 505, L99

\bibitem[{{Storchi-Bergmann} {et~al}\mbox{.}(2010){Storchi-Bergmann}, {Lopes},
  {McGregor}, {Riffel}, {Beck}, \& {Martini}}]{Storchi-Bergmann10}
{Storchi-Bergmann} T., {Lopes} R.~D.~S., {McGregor} P.~J., {Riffel} R.~A.,
  {Beck} T., {Martini} P., 2010, \mnras, 402, 819

\bibitem[{{Storchi-Bergmann} {et~al}\mbox{.}(2009){Storchi-Bergmann},
  {McGregor}, {Riffel}, {Sim{\~o}es Lopes}, {Beck}, \&
  {Dopita}}]{Storchi-Bergmann09}
{Storchi-Bergmann} T., {McGregor} P.~J., {Riffel} R.~A., {Sim{\~o}es Lopes} R.,
  {Beck} T., {Dopita} M., 2009, \mnras, 394, 1148

\bibitem[{{Storchi-Bergmann} {et~al}\mbox{.}(2012){Storchi-Bergmann}, {Riffel},
  {Riffel}, {Diniz}, {Borges Vale}, \& {McGregor}}]{Storchi-Bergmann12}
{Storchi-Bergmann} T., {Riffel} R.~A., {Riffel} R., {Diniz} M.~R., {Borges
  Vale} T., {McGregor} P.~J., 2012, \apj, 755, 87

\bibitem[{{Sturm} {et~al}\mbox{.}(2002){Sturm}, {Lutz}, {Verma}, {Netzer},
  {Sternberg}, {Moorwood}, {Oliva}, \& {Genzel}}]{Sturm02}
{Sturm} E., {Lutz} D., {Verma} A., {Netzer} H., {Sternberg} A., {Moorwood}
  A.~F.~M., {Oliva} E., {Genzel} R., 2002, \aap, 393, 821

\bibitem[{{Thompson} {et~al}\mbox{.}(2001){Thompson}, {Chary}, {Corbin}, \&
  {Epps}}]{Thompson01}
{Thompson} R.~I., {Chary} R.-R., {Corbin} M.~R., {Epps} H., 2001, \apjl, 558,
  L97

\bibitem[{{Unger} {et~al}\mbox{.}(1992){Unger}, {Lewis}, {Pedlar}, \&
  {Axon}}]{Unger92}
{Unger} S.~W., {Lewis} J.~R., {Pedlar} A., {Axon} D.~J., 1992, \mnras, 258, 371

\bibitem[{{Wang} {et~al}\mbox{.}(2012){Wang}, {Fabbiano}, {Karovska}, {Elvis}, 
  \& {Risaliti}}]{Wang12}
{Wang} J., {Fabbiano} G., {Karovska} M., {Elvis} M., {Risaliti} G., 2012, \apj,
  756, 180


\bibitem[{{Wilson} \& {Ulvestad}(1983)}]{Wilson83}
{Wilson} A.~S., {Ulvestad} J.~S., 1983, \apj, 275, 8

\end{thebibliography}
\end{document}